\documentclass[reprint, aip, footinbib,  letterpaper, superscriptaddress]{revtex4-2}

\usepackage[T1]{fontenc} 

\usepackage{amsmath,color}
\usepackage{amssymb}
\usepackage{amsfonts}
\usepackage{amsthm}
\usepackage{mathtools}

\usepackage{algorithm}
\usepackage{algpseudocode}
\usepackage{dsfont}

\usepackage{bbold}
\usepackage{chemformula}
\usepackage{siunitx}
\usepackage{pdfpages}
\usepackage{pgffor}

\DeclarePairedDelimiterX\braket[2]{\langle}{\rangle}{#1 \delimsize\vert #2}
\DeclarePairedDelimiterX\braket3[3]{\langle}{\rangle}{#1 \delimsize\vert #2 \delimsize\vert #3}

\usepackage{accents}

\usepackage{fancyvrb}

\newcommand{\vxi}{\boldsymbol{\xi}}

\newcommand{\tr}[1]{\text{Tr}\left[#1\right]}

\newcommand{\red}[1]{{\color{black} #1}}

\usepackage{adjustbox}

\makeatletter
\AtBeginDocument{\let\LS@rot\@undefined}
\makeatother

\begin{document}

       \title{Electronic Born--Oppenheimer Approximation in Nuclear-Electronic Orbital  Dynamics}
   
   \author{Tao E. Li}%
   \email{tao.li@yale.edu}
   \affiliation{Department of Chemistry, Yale University, New Haven, Connecticut 06520, USA}
   
   \author{Sharon Hammes-Schiffer}%
   \email{sharon.hammes-schiffer@yale.edu}
   \affiliation{Department of Chemistry, Yale University, New Haven, Connecticut 06520, USA}

   \begin{abstract}
   	Within the nuclear-electronic orbital (NEO) framework, the real-time NEO time-dependent density functional theory (RT-NEO-TDDFT) approach enables the simulation of coupled electronic-nuclear dynamics. In this approach, the electrons and quantum nuclei are propagated in time on the same footing.  A relatively small time step is required to propagate the much faster electronic dynamics, thereby prohibiting the simulation of long-time nuclear quantum dynamics. Herein, the electronic Born--Oppenheimer (BO) approximation within the NEO framework is presented. In this approach, the electronic density is quenched to the ground state at each time step, and the real-time nuclear quantum dynamics is propagated on an instantaneous electronic ground state defined by both the classical nuclear geometry and the nonequilibrium quantum nuclear density.  Because the electronic dynamics is no longer propagated, this approximation enables the use of an order-of-magnitude larger time step, thus greatly reducing the computational cost. Moreover, invoking the electronic BO approximation also fixes the unphysical asymmetric Rabi splitting observed in previous semiclassical RT-NEO-TDDFT simulations of vibrational polaritons even for small Rabi splitting, instead yielding a stable, symmetric Rabi splitting. For intramolecular proton transfer in malonaldehyde, both RT-NEO-Ehrenfest dynamics and its BO counterpart can describe proton delocalization during the real-time nuclear quantum dynamics. Thus, the BO RT-NEO approach provides the foundation for a wide range of chemical and biological applications.
   \end{abstract}

   \maketitle
   
   \section{Introduction}
   
   Simulating nuclear quantum dynamics is important for understanding a wide range of chemical and biological processes, such as vibrationally excited chemistry and proton transfer reactions.  A variety of theoretical methods for simulating nuclear quantum dynamics have been developed, including the linearized semiclassical initial value representation (LSC-IVR) method, \cite{Miller2001,Cotton2013} the ring polymer molecular dynamics method, \cite{Markland2018} the multiconfigurational time-dependent Hartree (MCTDH) method, \cite{Worth2008MCTDH} and the exact factorization approach. \cite{Abedi2010EF}  The nuclear-electronic orbital (NEO) method \cite{Webb2002,Pavosevic2020} is another approach for simulating nuclear quantum dynamics. \cite{Tao2021,Xu2022,Yu2022MSDFT,Zhao2020,Zhao2020JCP,Zhao2021JPCL,Li2022JCTCNEO} Within the NEO framework, both electrons and selected nuclei, usually the protons, are described by first-principles methods such as density functional theory (DFT),   while the remaining heavy nuclei are treated classically. As an extension of conventional electronic structure theory, the NEO method can be combined with conventional nonadiabatic dynamics approaches, such as Ehrenfest dynamics \cite{Li2005Eh,Isborn2007} and trajectory surface hopping, \cite{Tully1990} to simulate nonadiabatic dynamics on vibronic surfaces rather than electronic surfaces.
   
   A promising strategy for simulating nuclear quantum dynamics within the NEO framework is  real-time NEO time-dependent density functional theory (RT-NEO-TDDFT). \cite{Zhao2020} In this approach, the  quantum dynamics of the electrons and quantum protons are propagated in the time domain. As a real-time version of linear-response multicomponent TDDFT,\cite{Li1986TDDFTMulti,Marques2006TDDFT,Butriy2007,Yang2018} RT-NEO-TDDFT can capture the linear-response electronic and protonic excited-state spectra by Fourier transforming the real-time dipole signals. Moreover, this approach can also directly capture the  coupled electron-proton nonadiabatic dynamics. These calculations can be performed with fixed classical nuclei, or the classical nuclei can be propagated on the mean-field potential energy surface associated with the nonequilibrium electrons and quantum protons using the RT-NEO-Ehrenfest dynamics approach,\cite{Zhao2020JCP,Zhao2021JPCL} which captures the full nonadiabatic dynamics of the electrons, quantum protons, and classical nuclei. Another important extension of RT-NEO-TDDFT is semiclassical RT-NEO-TDDFT for polaritons, \cite{Li2022JCTCNEO} where the coupled dynamics between the NEO molecular subsystem and classical photon modes are propagated in the time domain. When the light-matter coupling is large enough, this approach can describe strong light-matter interactions under both electronic and vibrational strong couplings, thus potentially providing a powerful scheme for simulating polariton chemistry. \cite{Thomas2019_science,Li2022Review,Fregoni2022,Nagarajan2021}
   
   For these RT-NEO dynamics approaches, the electrons and quantum nuclei are propagated on the same footing. As a result, a very small time step ($\leq 0.01$ fs) is required for the time propagation due to the fast electronic dynamics. \cite{Zhao2020,Zhao2020JCP,Zhao2021JPCL,Li2022JCTCNEO} The requirement of a small time step may prohibit the study of long-time nuclear dynamics. When nuclear quantum dynamics on the electronic ground state is considered, such a small time step can be avoided by invoking the electronic Born--Oppenheimer (BO) approximation between the electrons and both quantum and classical nuclei. In this case, the electrons are quenched to the ground state for each time step of the dynamics, i.e., for each classical nuclear geometry and corresponding nonequilibrium quantum proton density. This electronic BO  approximation is different from the conventional BO approximation used in \textit{ab initio} molecular dynamics. In the conventional BO approximation, the electronic ground state is  determined solely by the geometry of the classical nuclei, whereas in the NEO electronic BO approximation, the electronic ground state is determined by the nonequilibrium proton density as well as the geometry of the classical nuclei. For a given classical nuclear geometry, the nonequilibrium proton density in a BO-RT-NEO dynamics simulation can be a mixture of vibrational states, thereby also influencing the electronic ground state. The BO-RT-NEO dynamics approach is also different from the constrained NEO (cNEO) dynamics method developed by Yang and coworkers. \cite{Xu2020} 
   The cNEO approach enables the inclusion of anharmonicity in vibrational spectra in a computationally efficient manner.\cite{Xu2022}  In addition to this capability, the  BO-RT-NEO approach also provides real-time dynamics associated with nonequilibrium proton densities.
   
   In this manuscript, we show that  invoking the BO approximation for RT-NEO dynamics allows an order-of-magnitude larger time step to be used during the time propagation compared to RT-NEO dynamics without the BO approximation. Thus, the BO approximation greatly reduces the computational cost, while producing nearly identical dynamics for the electronically adiabatic systems studied. Moreover, invoking the electronic BO approximation also overcomes a serious drawback in semiclassical RT-NEO dynamics for polaritons, namely the previously observed unphysical asymmetirc Rabi splitting under vibrational strong coupling even when the Rabi splitting is small. \cite{Li2022JCTCNEO} Herein we show that the BO-RT-NEO method produces symmetric Rabi splittings under these conditions. Lastly, using intramolecular proton transfer in malonaldehyde \cite{Baughcum1984,Baba1999,Barone1996,Tuckerman2001,Tautermann2002} as an example, we show that BO-RT-NEO-Ehrenfest dynamics can capture  proton delocalization associated with these types of proton transfer processes.

   \section{Theory}
   
   The equations of motion for RT-NEO dynamics, semiclassical RT-NEO dynamics for polaritons, and RT-NEO-Ehrenfest dynamics are summarized in Table \ref{table:eom}. A more detailed review of these approaches is given below.
   
   \begin{table*}
   	\caption{Equations of motion for different types of NEO dynamics.}
   	\begin{adjustbox}{width=\linewidth,center}
   		\label{table:eom}
   		\begin{tabular}{lccr}
   			\hline
   			\hline
   			& RT-NEO\cite{Zhao2020}\footnote{Here, RT-NEO refers to RT-NEO dynamics with fixed classical nuclei.}    & Semiclassical RT-NEO for polaritons \cite{Li2022JCTCNEO}  & RT-NEO-Ehrenfest\cite{Zhao2020JCP,Zhao2021JPCL}\\
   			\hline
   			electrons/non-BO & 	$i \frac{\partial}{\partial t} \mathbf{P}^{\text{e}}  = \left[\mathbf{F}^{\text{e}}, \mathbf{P}^{\text{e}} \right]$ & $i \frac{\partial}{\partial t} \mathbf{P}^{\text{e}}  = \left[\mathbf{F}^{\text{e}} + \sum_{k,\lambda}\varepsilon_{k,\lambda} q_{k,\lambda}\hat{\mu}_{\lambda}^{\text{e}}, \mathbf{P}^{\text{e}} \right]$ & $i \frac{\partial}{\partial t} \mathbf{P}^{\text{e}}  = \left[\mathbf{F}^{\text{e}}, \mathbf{P}^{\text{e}} \right]$ \\
   			electrons/BO       &   $\mathbf{P}^{\text{e}'} = \text{SCF}[\cdots]$ & $\mathbf{P}^{\text{e}'} = \text{SCF}[ \Re(\mathbf{P}^{\text{n}'}), \{\mathbf{R}_I\} ]$  & $\mathbf{P}^{\text{e}'} = \text{SCF}[\cdots]$   \\
   			\hline
   			quantum nuclei & 	$i \frac{\partial}{\partial t} \mathbf{P}^{\text{n}}  = \left[\mathbf{F}^{\text{n}}, \mathbf{P}^{\text{n}} \right]$ 
   			& $i \frac{\partial}{\partial t} \mathbf{P}^{\text{n}}  = \left[\mathbf{F}^{\text{n}}+ \sum_{k,\lambda}\varepsilon_{k,\lambda} q_{k,\lambda}\hat{\mu}_{\lambda}^{\text{n}}, \ \mathbf{P}^{\text{n}} \right]$ & $i \frac{\partial}{\partial t} \mathbf{P}^{\text{n}}  = \left[\mathbf{F}^{\text{n}}, \mathbf{P}^{\text{n}} \right]$\\
   			classical nuclei   & fixed &  fixed & $M_I \ddot{\mathbf{R}}_I = - \boldsymbol{\nabla}_I E$\\
   			photons & N/A & $\ddot{q}_{k,\lambda} = -\omega_{k,\lambda}^2 q_{k,\lambda} - \varepsilon_{k,\lambda} \mu_{\lambda} - \gamma_{\rm c} p_{k,\lambda}$ & N/A \\
   			\hline
   			\hline
   		\end{tabular}
   	\end{adjustbox}
   \end{table*}

   \subsection{RT-NEO dynamics with fixed classical nuclei}
   
   Within the framework of the RT-NEO approach with fixed classical nuclei, \cite{Zhao2020}  the dynamics of both electrons (assuming closed-shell) and quantum nuclei are propagated by the following von Neumann equations
   \begin{subequations}\label{eq:rt_NEO}
   	\begin{align}
   		\label{eq:rt_NEO-e}
   		i \frac{\partial}{\partial t} \mathbf{P}^{\text{e}}(t) & = \left[\mathbf{F}^{\text{e}}(t), \ \mathbf{P}^{\text{e}}(t) \right] \\
   		\label{eq:rt_NEO-p}
   		i \frac{\partial}{\partial t} \mathbf{P}^{\text{n}}(t) & = \left[ \mathbf{F}^{\text{n}}(t),\ \mathbf{P}^{\text{n}}(t) \right ]
   	\end{align}
   \end{subequations}
   Here, the density matrices are defined as $\mathbf{P}^{\text{e}} =  \mathbf{C}^{\text{e}}\mathbf{C}^{\text{e}\dagger}$ and  $\mathbf{P}^{\text{n}} = \mathbf{C}^{\text{n}}\mathbf{C}^{\text{n}\dagger}$, where $\mathbf{C}^{\text{e}}$ (or $\mathbf{C}^{\text{n}}$) denotes the coefficient matrix of the electronic (or nuclear) wavefunction in the orthogonal atomic orbital basis. The transformation between the density matrices in the orthogonal (labeled without prime) and non-orthogonal (labeled with prime) atomic basis is governed by	 
   \begin{equation}\label{eq:Pe_transform}
   	\begin{aligned}
   		\mathbf{P}^{\text{e}} = [\mathbf{S}^{\text{e}}]^{1/2} \mathbf{P}^{\text{e}'}[\mathbf{S}^{\text{e}}]^{1/2} 
   	\end{aligned}
   \end{equation}
   where $\mathbf{S}^{\text{e}}$ is the electronic  overlap matrix.
   Similarly, in Eq. \eqref{eq:rt_NEO-e}, the Kohn--Sham matrices $\mathbf{F}^{\text{e}}$ in the orthogonal atomic orbital basis are defined as
   \begin{equation}\label{eq:Fe_transform}
   	\begin{aligned}
   		\mathbf{F}^{\text{e}} = [\mathbf{S}^{\text{e}}]^{-1/2} \mathbf{F}^{\text{e}'}[\mathbf{S}^{\text{e}}]^{-1/2} 
   	\end{aligned}
   \end{equation}
   Here, $\mathbf{F}^{\text{e}'}$ denotes the Kohn--Sham matrix for the electrons in the non-orthogonal atomic orbital basis. The analogs to Eqs. \eqref{eq:Pe_transform} and \eqref{eq:Fe_transform} for the quantum nuclei are identical with the superscript e replaced by n.
   
   According to NEO-DFT,\cite{Pak2007,Chakraborty2008}  $\mathbf{F}^{\text{e}'}$ and $\mathbf{F}^{\text{n}'}$ are defined as
   \begin{subequations}\label{eq:KS_matrix}
   	\begin{equation}
   		\begin{aligned}
   			\label{eq:KS_matrix-e}
   			\mathbf{F}^{\text{e}'}(t) &= \mathbf{H}_{\text{core}}^{\text{e}'} + \mathbf{J}^{\text{ee}'}(\mathbf{P}^{\text{e}'}(t)) + \mathbf{V}_{\text{xc}}^{\text{e}'}(\mathbf{P}^{\text{e}'}(t)) \\ &
   			-\mathbf{J}^{\text{en}'}(\mathbf{P}^{\text{n}'}(t))
   			+\mathbf{V}_{\text{c}}^{\text{en}'}(\mathbf{P}^{\text{e}'}(t), \mathbf{P}^{\text{n}'}(t))
   		\end{aligned}
   	\end{equation}
   	
   	\begin{equation}
   		\begin{aligned}
   			\label{eq:KS_matrix-p}
   			\mathbf{F}^{\text{n}'}(t) &= \mathbf{H}_{\text{core}}^{\text{n}'} + \mathbf{J}^{\text{nn}'}(\mathbf{P}^{\text{n}'}(t)) + \mathbf{V}_{\text{xc}}^{\text{n}'}(\mathbf{P}^{\text{n}'}(t)) \\ &
   			-\mathbf{J}^{\text{ne}'}(\mathbf{P}^{\text{e}'}(t))
   			+\mathbf{V}_{\text{c}}^{\text{ne}'}(\mathbf{P}^{\text{n}'}(t), \mathbf{P}^{\text{e}'}(t))
   		\end{aligned}
   	\end{equation}
   \end{subequations}
   In Eq. \eqref{eq:KS_matrix}, $\mathbf{H}_{\text{core}}^{\text{e}'}$ (or $\mathbf{H}_{\text{core}}^{\text{n}'}$) denotes the core Hamiltonian, which includes the kinetic energy and the Coulomb interaction between the electrons (or quantum nuclei) and the classical nuclei;  $\mathbf{J}^{\text{ee}'}$ (or $\mathbf{J}^{\text{nn}'}$) denotes the Coulomb interactions among the electrons (or quantum nuclei);  $\mathbf{V}_{\text{xc}}^{\text{e}'}$ (or $\mathbf{V}_{\text{xc}}^{\text{n}'}$) denotes the exchange-correlation potential for the electrons (or quantum nuclei);
   $\mathbf{J}^{\text{en}'}$ (or $\mathbf{J}^{\text{ne}'}$) denotes the Coulomb interaction between the electrons and quantum nuclei; and $\mathbf{V}_{\text{c}}^{\text{en}'}$ (or $\mathbf{V}_{\text{c}}^{\text{ne}'}$) denotes the correlation potential between the electrons and quantum nuclei. In the Hartree--Fock limit, $\mathbf{V}_{\text{c}}^{\text{ne}'} = \mathbf{V}_{\text{c}}^{\text{en}'} = \mathbf{0}$, and
   $\mathbf{V}_{\text{xc}}^{\text{e}'}$ (or $\mathbf{V}_{\text{xc}}^{\text{n}'}$) becomes the Hartree--Fock exchange term for electrons (or quantum nuclei).
   \red{
   	Note that the nuclear-nuclear exchange-correlation potential $\mathbf{V}_{\text{xc}}^{\text{n}'}$ terms are many orders of magnitude smaller than the analogous electronic terms $\mathbf{V}_{\text{xc}}^{\text{e}'}$ for molecular systems, where the proton orbitals are localized. \cite{Pavosevic2020} Hence, typically the nuclear-nuclear Hartree--Fock exchange terms are included to avoid self-interaction error, but the nuclear-nuclear correlation terms are neglected in NEO calculations.} Similar to most RT-TDDFT implementations,\cite{Goings2018,Isborn2008} the adiabatic approximation is  invoked, and the above functionals depend locally on time.
   
   \subsection{Semiclassical RT-NEO dynamics for polaritons}
   
   Beyond RT-NEO with fixed classical nuclei, the semiclassical RT-NEO approach for polaritons \cite{Li2022JCTCNEO} can describe strong light-matter interactions between cavity photon modes and molecules \cite{Galego2019,Campos-Gonzalez-Angulo2019,LiHuo2021,Schafer2021,Flick2017,Rosenzweig2022,Riso2022}. Within this approach, the cavity photons are propagated classically: 
   \begin{subequations}\label{eq:cavity_dynamics}
   	\begin{align}
   		\dot{q}_{k,\lambda} &= p_{k,\lambda} \\
   		\dot{p}_{k,\lambda} &= -\omega_{k,\lambda}^2 q_{k,\lambda} - \varepsilon_{k,\lambda} \mu_{\lambda} - \gamma_{\rm c} p_{k,\lambda}
   	\end{align}
   \end{subequations}
   Here, ${q}_{k,\lambda}$, ${p}_{k,\lambda}$, and ${\omega}_{k,\lambda}$ denote the position, momentum, and frequency of the cavity photon mode characterized by the wave vector $k = |\mathbf{k}|$ and polarization unit vector $\vxi_\lambda$, where $\mathbf{k}\cdot \vxi_{\lambda} = 0$ (e.g., if the $\mathbf{k}$ direction is $z$, $\lambda$ can be $x$ or $y$); $\varepsilon_{k,\lambda}$ denotes the light-matter coupling; $\mu_{\lambda}$ denotes the dipole moment of the molecule along the direction of $\vxi_{\lambda}$; $\gamma_{\rm c}$ denotes the cavity loss rate. In practice, when calculating $\mu_{\lambda}(t)$, we subtract the permanent dipole contribution, i.e., $\mu_{\lambda}(t) = 2\tr{\mathbf{P}^{\rm e}(t)\hat{\mu}^{\rm e}_{\lambda}} + \tr{\mathbf{P}^{\rm n}(t)\hat{\mu}^{\rm n}_{\lambda}} - 2\tr{\mathbf{P}^{\rm e}(0)\hat{\mu}^{\rm e}_{\lambda}} - \tr{\mathbf{P}^{\rm n}(0)\hat{\mu}^{\rm n}_{\lambda}}$, so at time $t = 0$, $q_{k,\lambda} = p_{k,\lambda} = 0$ always represents the photonic ground state. \cite{Li2022JCTCNEO} Here, $\hat{\mu}_{\lambda}^{\text{e}}$ (or $\hat{\mu}_{\lambda}^{\text{n}}$) denotes the dipole matrix of the electrons (or quantum nuclei) projected along the direction of $\vxi_{\lambda}$ in the orthogonal atomic orbital basis, and the prefactor 2 in the electronic dipole moment is included because of the restricted Kohn--Sham calculation.
   
   Due to the interaction with cavity photons, the dynamics of the electrons and quantum nuclei become
   \begin{subequations}\label{eq:semiclassical_rt_NEO}
   	\begin{align}
   		\label{eq:semiclassical_rt_NEO-e}
   		i \frac{\partial}{\partial t} \mathbf{P}^{\text{e}}(t) & = \left[\mathbf{F}^{\text{e}}(t) + \sum_{k,\lambda}\varepsilon_{k,\lambda} q_{k,\lambda}\hat{\mu}_{\lambda}^{\text{e}}, \ \mathbf{P}^{\text{e}}(t) \right] \\
   		\label{eq:semiclassical_rt_NEO-p}
   		i \frac{\partial}{\partial t} \mathbf{P}^{\text{n}}(t) & = \left[ \mathbf{F}^{\text{n}}(t) + \sum_{k,\lambda}\varepsilon_{k,\lambda} q_{k,\lambda}\hat{\mu}_{\lambda}^{\text{n}},\ \mathbf{P}^{\text{n}}(t) \right ]
   	\end{align}
   \end{subequations}
   In Eqs. \eqref{eq:cavity_dynamics} and \eqref{eq:semiclassical_rt_NEO}, although many cavity modes indexed by $k,\lambda$ have been considered, in the simulation below, for simplicity, we will  take into account only one cavity mode polarized along the $x$-direction. 
   
   Similar to RT-NEO dynamics with fixed classical nuclei, here the classical nuclei are also assumed to be  fixed. Eqs. \eqref{eq:cavity_dynamics} and \eqref{eq:semiclassical_rt_NEO} can be further combined with a mean-field propagation of the classical nuclei via Ehrenfest dynamics, thus providing a full dynamics scheme for polariton chemistry applications. Because this extension is beyond the scope of this manuscript, we will report this development elsewhere.

   \subsection{RT-NEO-Ehrenfest dynamics}
   
   The RT-NEO-Ehrenfest dynamics\cite{Zhao2020JCP,Zhao2021JPCL} method combines the real-time dynamics of the electrons and quantum nuclei and the mean-field motion of the classical nuclei. In this approach, the electrons and quantum nuclei are propagated according to Eq. \eqref{eq:rt_NEO},  and the remaining nuclei are propagated classically by the following equations of motion:
   \begin{subequations}
   	\begin{align}
   		\dot{\mathbf{R}}_I &= \frac{\mathbf{P}_I}{M_I} \\
   		\dot{\mathbf{P}}_I &= - \boldsymbol{\nabla}_I E[\mathbf{P}^{\text{e}'}(t), \mathbf{P}^{\text{n}'}(t), \{\mathbf{R}_I\}]
   	\end{align}
   \end{subequations}
   Here, $\mathbf{R}_I$, $\mathbf{P}_I$, and $M_I$ denote the position, momentum, and mass of the $I$-th classical nucleus;  the total energy of the molecular system $E[\mathbf{P}^{\text{e}'}(t), \mathbf{P}^{\text{n}'}(t), \{\mathbf{R}_I\}]$ is a function of the nonequilibrium  densities of the electrons and quantum nuclei ($\mathbf{P}^{\text{e}'}(t)$ and $\mathbf{P}^{\text{n}'}(t)$) as well as the positions of all classical nuclei $\mathbf{R}_I$. Ref. \onlinecite{Zhao2020JCP} provides the explicit form of the Ehrenfest gradients $\boldsymbol{\nabla}_I E$.
   
   When using RT-NEO-Ehrenfest dynamics for describing proton transfer, a reasonable choice for treating the proton basis function centers is to use a large proton basis set including several different fixed proton basis (FPB) function centers spanning the region sampled by the transferring proton. \cite{Zhao2020} Another choice is the traveling proton basis (TPB) approach,\cite{Zhao2020JCP,Zhao2021JPCL}  in which the proton basis function centers are allowed to move semiclassically along the proton transfer trajectory. Because this TPB approach is a semiclassical approximation of the FPB approach, we will focus on the FPB approach in this manuscript. However, the RT-NEO-Ehrenfest dynamics simulations performed herein can also be performed with the TPB approach in a straightforward manner.  Moreover, as mentioned above, the RT-NEO-Ehrenfest approach can also be used in conjunction with semiclassical RT-NEO dynamics for polaritons.

   \subsection{Electronic BO Approximation}
   
   When the electronic BO approximation is applied, the protonic dynamics is still propagated by Eq. \eqref{eq:rt_NEO-p} or Eq. \eqref{eq:semiclassical_rt_NEO-p}. For the electrons, at each time step, the electronic density matrix is quenched to the ground state by solving the electronic self-consistent field (SCF) equation: 
   \begin{equation}\label{eq:e_scf}
   	\mathbf{P}^{\text{e}'}(t) = \text{SCF}[ \Re(\mathbf{P}^{\text{n}'}(t)), \{\mathbf{R}_I(t)\} ]
   \end{equation}
   Here, $\Re(\mathbf{P}^{\text{n}'}(t))$ denotes the real component of the protonic density matrix in the non-orthogonal atomic orbital basis. Because the converged $\mathbf{P}^{\text{e}'}$ and electronic energy are real-valued, the imaginary component of $\mathbf{P}^{\text{n}'}$ does not need to be included in the electronic Kohn-Sham matrix in   Eq. \eqref{eq:KS_matrix-e} for this SCF procedure. Hence, only $\Re(\mathbf{P}^{\text{n}'}(t))$ is used to solve the electronic SCF equation. More specifically, when solving Eq. \eqref{eq:e_scf}, we iteratively find the converged electronic density satisfying the following Hartee--Fock--Roothaan equation:
   \begin{equation}
   	\mathbf{F}^{\rm e'} \mathbf{C}^{\rm e'} = \mathbf{S}^{\rm e}\mathbf{C}^{\rm e'} \boldsymbol{\varepsilon}^{\rm e}
   \end{equation}    
   where $\boldsymbol{\varepsilon}^{\rm e}$ is the orbital energy matrix and $\mathbf{F}^{\rm e'}$ and $\mathbf{C}^{\rm e'} $ have been defined above. Here, $\mathbf{F}^{\rm e'}$ is a function of $\mathbf{P}^{\rm e'}$ (or $\mathbf{C}^{\rm e'}$), 
   $\Re(\mathbf{P}^{\text{n}'})$, and $\{\mathbf{R}_I\}$.
   
   \section{Simulation Details}
   
   All the above approaches have been implemented in a developer version of Q-Chem. \cite{Epifanovsky2021} The initial molecular geometries for the calculations below are given in the Supplementary Material. The electrons and quantum nuclei were propagated by a modified midpoint unitary transform time-propagation scheme algorithm. \cite{Goings2018,Li2005} During time propagation, an additional predictor-corrector procedure \cite{DeSantis2020} was used to control the growth of numerical error in the electronic and nuclear quantum dynamics. The velocity Verlet algorithm was used to propagate the classical nuclei and cavity modes. The step-by-step algorithms of the above approaches are provided in the Supplementary Material. 
   
   The RT-NEO method with fixed classical nuclei was applied to a single HCN molecule. The B3LYP functional \cite{Lee1988,Becke1988,Becke1998} was used for electron-electron exchange-correlation and the epc17-2 functional \cite{Brorsen2017,Yang2017} was used for electron-proton correlation. For the electronic basis, the cc-pVDZ electronic basis set \cite{Dunning1989} was used for the heavy nuclei and the cc-pV5Z electronic basis set was used for the proton; for the protonic basis, the PB4-F2 proton basis set \cite{Yu2020NEOBasis} was used. The initial densities for the electrons and quantum proton were obtained from the SCF ground state NEO-DFT solution\cite{Pak2007,Chakraborty2008} with a tight energy convergence criterion of $10^{-12}$ a.u. During the real-time propagation, at time $t = 0$ a delta pulse was used to \red{perturb} the protonic Fock matrix as $\mathbf{F}^{\rm n'} + \mathbf{E}\cdot \boldsymbol{\mu}^{\rm n '}$, where $\mathbf{E} = (E_0, E_0, E_0)$ and $\boldsymbol{\mu}^{\rm n '}$ denotes the protonic dipole moment matrix vector in the non-orthogonal atomic orbital basis.  Because we will compare the performance of calculations with different time steps $\Delta t_{\rm q}$,  $E_0\Delta t_{\rm q} = 4\times 10^{-4}$ a.u. is always assumed. This restriction ensures that simulations with different time steps will produce dipole signals with the same amplitude. As we will compare the performance with and without the electronic BO approximation, the delta pulse was not applied to the electronic subsystem. 
   
   The same HCN molecule was also used in the semiclassical RT-NEO calculations for polaritons. A single cavity mode polarized along the $x$ direction was resonantly coupled to the $\ch{C-H}$ stretch mode of the molecule at $\omega_{\rm c} = 3685$ cm$^{-1}$ with light-matter coupling $\varepsilon = 6\times 10^{-4}$ a.u. The initial conditions and computational methods for \ch{HCN} were the same as those described for the RT-NEO calculations in free space. The initial condition for the cavity mode was set as $p_{\rm c}(0) = 0$ and $q_{\rm c}(0) = 0.1$ a.u., and no delta pulse was applied to the molecular subsystem. Because the position of the cavity mode was displaced to 0.1 a.u., in later times the excess energy in the cavity mode transferred to the $\ch{C-H}$ stretch mode and generated real-time Rabi oscillations.  The cavity loss rate was assumed to be  $\gamma_{\rm c} = 0$.
   
   The RT-NEO-Ehrenfest method was applied to intramolecular proton transfer reaction in malonaldehyde with the transferring proton treated quantum mechanically. The molecule was described at the B3LYP/epc17-2/cc-pVDZ/PB4-F2 level. The initial electronic and quantum protonic densities were obtained from the NEO-DFT ground state solution with an energy convergence criterion of $10^{-9}$ a.u. The initial velocities of the classical nuclei were set to zero. Because these classical nuclei were chosen to start out in a symmetric configuration, whereas the equilibrium geometry for the classical nuclei is asymmetric, the classical nuclei experienced forces directed toward the asymmetric relaxed geometry, thus driving proton transfer. For the quantum proton, three fixed proton basis function centers were used, and each center contained a PB4-F2 proton basis set and a cc-pVDZ electronic basis set. These three proton basis function centers were chosen to be near the donor oxygen atom (O$_{\rm D}$), the acceptor oxygen atom (O$_{\rm A}$), and the midpoint between the two centers. The quantum proton position is defined as the expectation value of the proton position operator. For RT-NEO-Ehrenfest dynamics without the electronic BO approximation, by default we set the time step for the electronic and protonic quantum dynamics as $\Delta t_{\rm q} = 0.010$ fs, and the classical nuclear gradients were evaluated every 10 time steps of the quantum dynamics. Under the electronic BO approximation, because a 10-fold larger time step was used for the protonic quantum dynamics, the nuclear gradients were evaluated at every time step of the quantum dynamics.

   \section{Results and Discussion}

   \subsection{RT-NEO vs BO-RT-NEO dynamics with fixed classical nuclei}
\begin{figure*}
   	\centering
   	\includegraphics[width=1.0\linewidth]{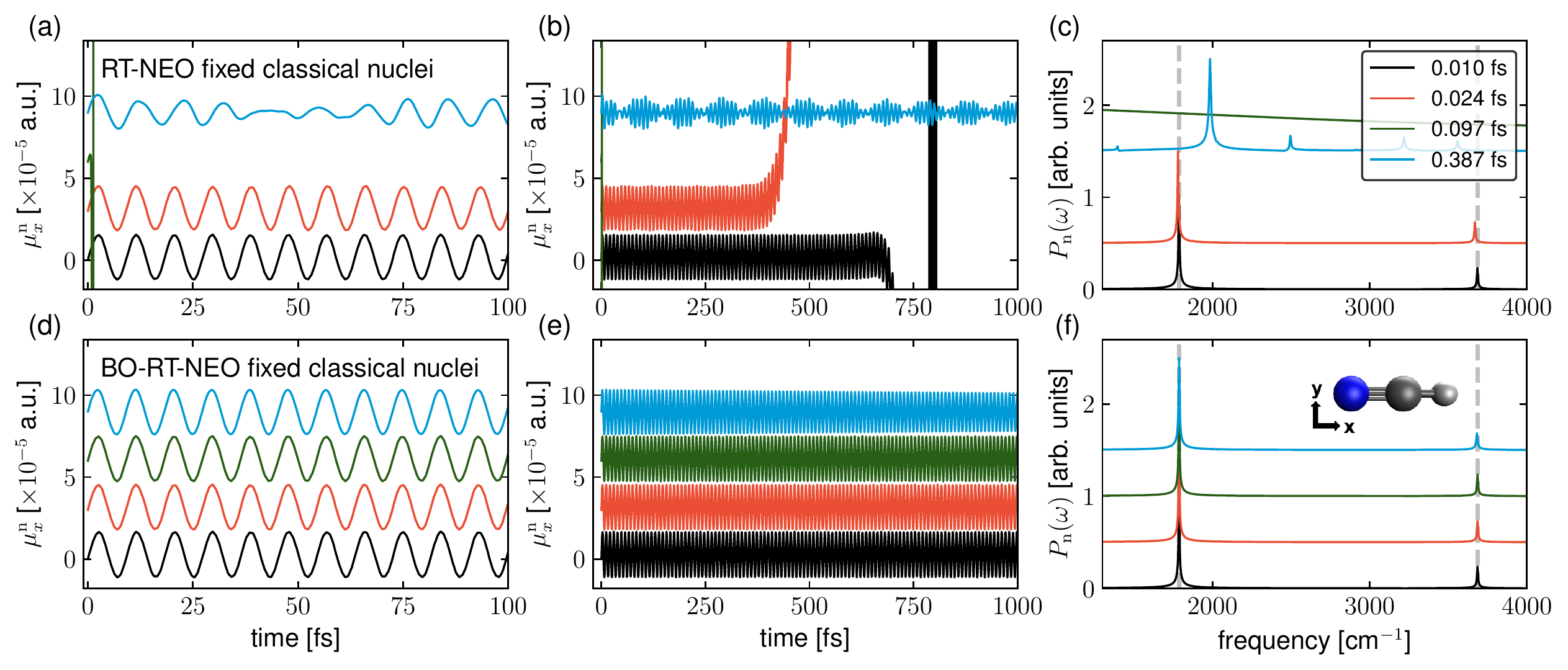}
   	\caption{(a) RT-NEO-TDDFT dynamics  of the $x$-component of the \ch{HCN} nuclear dipole moment, $\mu_x^{\text{n}}(t)$, in vacuum with fixed classical nuclei. The quantum proton is perturbed by a delta pulse at $t = 0$ fs. Simulation results with different time steps are compared: $\Delta t_{\rm q} =$ 0.010 fs (black), 0.024 fs (red), 0.097 fs (dark green), and 0.387 fs (blue).  \red{(b) Same trajectories as (a) but extended to 1 ps. (c)}  Corresponding nuclear dipole power spectrum, $P_{\rm n}(\omega)$, as well as the linear-response NEO-TDDFT  vibrationally excited state transition frequencies (vertical  gray dashed lines). \red{(d)-(f) Analogous plots as (a)-(c)} except that the BO-RT-NEO-TDDFT approach is used. \red{For all plots, signals with the three larger time steps are each shifted up by a constant for better visualization. Although the RT-NEO approach requires a relatively small $\Delta t_{\rm q}$ to produce accurate proton frequencies, the BO-RT-NEO approach can still produce accurate proton frequencies using a 16-fold larger $\Delta t_{\rm q}$. Without the electronic BO approximation, using relatively large time steps such as  $\Delta t_{\rm q}=0.097$ fs (as well as 0.194 fs, not shown here) can lead to divergences in both the time and frequency domains, as indicated by the vertical dark green lines at early times in (a) and (b) and the horizontal dark green line in (c). The larger time step of 0.387 fs does not lead to such divergences but still leads to a qualitatively inaccurate vibrational spectrum.}
   	}
   	\label{fig:mun_linear_response}
   \end{figure*}

   To compare the performance of RT-NEO and BO-RT-NEO with fixed classical nuclei, we applied both approaches to a single \ch{HCN} molecule oriented along the $x$-axis.  Fig. \ref{fig:mun_linear_response}a shows the RT-NEO-TDDFT dynamics of the $x$-component of the \ch{HCN} nuclear dipole moment, $\mu_x^{\text{n}}(t) = \tr{\mathbf{P^{\rm n}}(t)\hat{\mu}_x^{\rm n}}$,  when the ground-state proton density is perturbed by a delta pulse at time $t = 0$ fs. As shown in this figure, \red{within 100 fs, the RT-NEO-TDDFT approach yields similar dipole oscillations for any time step $\Delta t_{\rm q} \leq 0.024$ fs, whereas a larger time step $\Delta t_{\rm q} \geq 0.097$ fs leads to either divergence or inaccurate oscillations. When the long-time dynamics of $\mu_x^{\text{n}}(t)$ is considered, as shown in Fig. \ref{fig:mun_linear_response}b for 1 ps, long-time divergence is observed even for a relatively small time step $\Delta t_{\rm q} = 0.010$ fs. Interestingly, the dipole oscillations using a very large $\Delta t_{\rm q} = 0.387$ fs become stable, presumably due to the use of the predictor-corrector procedure \cite{DeSantis2020} in the algorithm, which may help control the growth of numerical error in some cases. Despite the numerical stability, however, the predicted vibrational spectrum is qualitatively incorrect for this large time step.} 
   
   Fig. \ref{fig:mun_linear_response}\red{c} plots the corresponding power spectrum of the dipole signal in Fig. \ref{fig:mun_linear_response}\red{b} calculated by the following Fourier transform:
   \begin{equation}\label{eq:Pn_spectrum}
   	P_{\rm n}(\omega) = \sum_{i=x,y,z}\left | \mathcal{F}\left[\mu^{\rm n}_{i}(t) e^{-\gamma t}\right] \right |
   \end{equation}
   Here, for a better visualization of the spectrum, a small, artificial damping term $e^{-\gamma t}$ (with $\gamma = 10^{-5}$ a.u.)  provides  a small linewidth of 13.8 cm$^{-1}$ for the peaks in the frequency domain.  The Pad\'e approximation of the Fourier transform \cite{Bruner2016,Goings2018} is used in Eq. \eqref{eq:Pn_spectrum} for better frequency resolution. \red{Consistent spectra are obtained only when $\Delta t_{\rm q} \leq 0.024$ fs}: the peak at 1787 cm$^{-1}$ is the \ch{C-H} bend mode, and  the peak at 3685 cm$^{-1}$ is the \ch{C-H} stretch mode. \red{Note that these vibrational frequencies differ significantly from the experimental values, \cite{Yang2018,Culpitt2019JCP} mainly due to the use of insufficient electronic and protonic basis sets. As shown previously, quantitative accuracy of NEO-TDDFT frequencies can be achieved by increasing the size of the electronic and protonic basis sets.\cite{Culpitt2019JCP}} When $\Delta t_{\rm q} \red{= 0.010}$ fs, the RT-NEO-TDDFT frequencies agree well with the linear-response NEO-TDDFT results, as denoted by the vertical gray dashed lines in the spectrum. \red{When $\Delta t_{\rm q} = 0.024$ fs, the RT-NEO-TDDFT frequencies start to deviate from the linear-response results.} Note that the \red{real-time results} also predict the same absorption intensities as the linear-response results, as shown in Fig. S1 of the Supplementary Material.

   Figures \ref{fig:mun_linear_response}\red{d-f} show the analogous results obtained with the BO-RT-NEO-TDDFT approach instead of  the RT-NEO-TDDFT approach. Here, all the time steps \red{$0.010 \leq \Delta t_{\rm q} \leq 0.387$} fs  yield the same real-time dynamics and spectrum.  \red{When Fig. \ref{fig:mun_linear_response}f is compared to Fig. \ref{fig:mun_linear_response}c for $\Delta t_{\rm q} = 0.010$ fs, the unchanged protonic spectrum after applying the electronic BO approximation suggests that electronic excited states do not contribute significantly to the low-frequency protonic excited-state solutions, in agreement with linear-response NEO-TDDFT results (Table S1 in the Supplementary Material), and that energy transfer between the electronic and protonic degrees of freedom along the RT-NEO-TDDFT trajectory is negligible. In the future, it would be interesting to investigate molecular systems with significant energy transfer between protonic and electronic excited states using existing analysis tools. \cite{Shao2020PCCP}} 
   
   We find that the BO-RT-NEO-TDDFT approach produces reliable \red{frequencies} with \red{a 16-fold} larger time step than that required for the RT-NEO-TDDFT approach.  Because the electronic density needs to be quenched to the ground state by solving the electronic SCF equation (see Eq. \eqref{eq:e_scf}) for each time step of the electronic BO dynamics, each time step for BO-RT-NEO-TDDFT is more expensive than a time step for RT-NEO-TDDFT. In practice, we find that for the calculations in Fig. \ref{fig:mun_linear_response}, the computational cost of BO-RT-NEO-TDDFT per time step is \red{about twice that} of RT-NEO-TDDFT. Hence, overall, invoking the electronic BO approximation can accelerate the calculation of the proton vibrational spectrum by a factor of \red{$\gtrsim 8$. When the long-time quantum nuclear dynamics is considered, as shown in Fig. \ref{fig:mun_linear_response}e and Fig. \ref{fig:mun_linear_response}b, invoking the electronic BO approximation can also ensure long-time stability for a wide range of time steps.} 
   
   \subsection{Semiclassical RT-NEO vs semiclassical BO-RT-NEO dynamics for polaritons}

   \begin{figure*}
   	\centering
   	\includegraphics[width=0.9\linewidth]{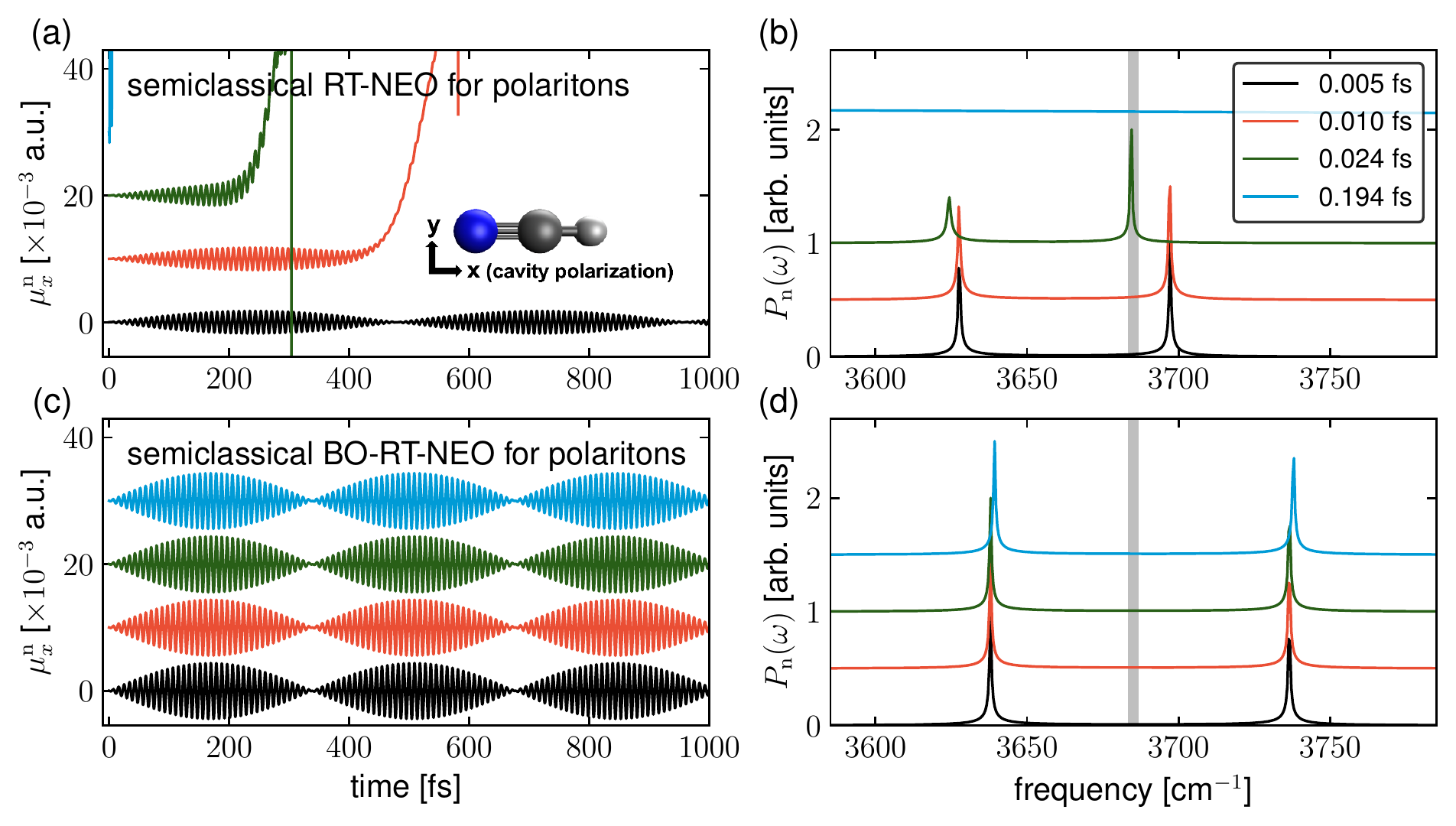}
   	\caption{ (a) Semiclassical RT-NEO-TDDFT dynamics of $\mu_x^{\text{n}}(t)$ and (b) the corresponding power spectrum  when a single \ch{HCN} molecule is coupled to an $x$-polarized cavity mode with $\omega_{\rm c} = 3685$ cm$^{-1}$ (the vertical thick gray line in (b)) and coupling strength $\varepsilon = 6\times 10^{-4}$ a.u. Simulation results with different time steps  are compared: \red{$\Delta t_{\rm q} =$ 0.005 fs (black), 0.010 fs (red), 0.024 fs (dark green), and 0.194 fs (blue). Signals with the three larger time steps are each shifted up by a constant for better visualization.} (c,d) Analogous plots as (a,b) except that the semiclassical BO-RT-NEO-TDDFT approach is used. Without the electronic BO approximation, using time steps greater than $\Delta t_{\rm q}=0.005$ fs can lead to divergences in both the time and frequency domains, as indicated by the vertical blue, dark green, and red lines in (a). Invoking the BO approximation not only generates stable long-time simulation results when a larger $\Delta t_{\rm q}$ is used, but also removes the unphysical asymmetry in the Rabi splitting observed by the semiclassical RT-NEO approach. 
   	}
   	\label{fig:mun_incav}
   \end{figure*}
   
   Beyond the conventional RT-NEO-TDDFT dynamics with fixed classical nuclei, when the molecular system is coupled to a classical cavity photon mode, the semiclassical RT-NEO-TDDFT approach can provide a unified description of vibrational and electronic strong couplings. \cite{Li2022JCTCNEO} Fig. \ref{fig:mun_incav}a shows  $\mu_x^{\text{n}}(t)$ for the $x$-oriented \ch{HCN} molecule under vibrational strong coupling  when the molecule is  resonantly coupled to an $x$-polarized cavity mode with the cavity frequency the same as the \ch{C-H} stretch mode ($\omega_{\rm c} = 3685$ cm$^{-1}$). 
   After an initial perturbation of the cavity mode, the coherent energy transfer between the cavity mode and the molecule leads to vibrational Rabi oscillations in the time domain. However, this behavior is only reliably captured when the simulation time step is $\Delta t_{\rm q} = 0.005$ fs (black line), whereas using a larger time step $\Delta t_{\rm q} \geq 0.010$ fs (red, green, and blue lines) leads to divergent time-domain dynamics.
   
   In the frequency domain, the vibrational polariton spectrum can be calculated from the power spectrum of $\mu^{\rm n}_{x}(t)$ using Eq. \eqref{eq:Pn_spectrum}.  Although the time-domain dynamics is converged only when $\Delta t_{\rm q} =$ 0.005 fs \red{(black line)}, in the frequency domain the results with $\Delta t_{\rm q} =$ 0.005 fs \red{(black line)} and $\Delta t_{\rm q} =$ 0.010 fs \red{(red line)} are virtually identical, presumably because the $\Delta t_{\rm q} =$ 0.010 fs real-time dipole dynamics is stable for a relatively long time (up to $t \sim 300$ fs).  For the converged spectrum (black or red line), the two polariton peaks are asymmetric with respect to the resonance frequency ($\omega_{\rm c} = 3685$ cm$^{-1}$, vertical thick gray line). Because the Rabi splitting between the two polaritons is very small (within 100 cm$^{-1}$) compared to the resonance frequency $\omega_{\rm c}$, the polariton peaks are expected to be symmetric with respect to the resonance frequency. \cite{FriskKockum2019,Li2020Water,George2015} Hence, this observed asymmetry is unphysical, reflecting a  limitation of the semiclassical RT-NEO approach when describing vibrational strong coupling.
   
   Figs. \ref{fig:mun_incav}c and d show  the dipole dynamics and spectra obtained with the electronic BO approximation under the same conditions. Comparing Fig. \ref{fig:mun_incav}c and Fig. \ref{fig:mun_incav}a, we find that invoking the electronic BO approximation captures stable real-time Rabi oscillations even with a 40-fold larger time step ($\Delta t_{\rm q} = 0.194$ fs,  \red{blue} line) than that required for the semiclassical RT-NEO approach ($\Delta t_{\rm q} = 0.005$ fs, \red{black} line). Moreover, in the frequency domain (Fig. \ref{fig:mun_incav}d), the two polariton peaks are  symmetric with respect to the resonance frequency ($\omega_{\rm c} = 3685$ cm$^{-1}$, vertical thick gray line), representing a significant advantage of the BO-RT-NEO-TDDFT approach for simulating molecular polaritons.
   
   The comparison between these two approaches allows us to understand why semiclassical RT-NEO-TDDFT without the electronic BO approximation predicts  asymmetric polaritons even when the Rabi splitting is very small. Comparing Fig. \ref{fig:mun_incav}a and Fig. \ref{fig:mun_incav}c, the nuclear dipole signals without the electronic BO approximation have smaller oscillation amplitudes, suggesting that in this case the cavity energy is also transferred to the electronic degrees of freedom in the molecular system, as the cavity mode is coupled to both the electronic and nuclear Kohn--Sham matrices (see Eq. \eqref{eq:semiclassical_rt_NEO}). Due to this energy transfer, the higher-energy electronic excitations  influence the vibrational polariton spectrum, leading to the asymmetry in the vibrational Rabi splitting. 
   
   This argument can also be understood by a simple three-state model:
   \begin{equation}
   	H = 
   	\begin{pmatrix}
   		\omega_{\rm c} & g_{\rm v} & g_{\rm e} \\
   		g_{\rm v} & \omega_{\rm v} & 0 \\
   		g_{\rm e} & 0 & \omega_{\rm e}
   	\end{pmatrix}
   \end{equation}
   where $g_{\rm v}$ ($g_{\rm e}$) denotes the coupling between the cavity mode and the vibrational (electronic) transition of frequency $\omega_{\rm v}$ ($\omega_{\rm e}$)\red{, which is proportional to the vibrational (electronic) transition dipole moment of the molecule.}
   In the absence of the electronic transition (i.e., let $g_{\rm e} = 0$), at  resonance condition $\omega_{\rm c} = \omega_{\rm v} = \omega_0$, the two polariton frequencies are
   \begin{equation}
   	\omega_{\pm} = \omega_{0} \pm g_{\rm v}
   \end{equation}
   which is symmetric with respect to $\omega_0$.
   Due to the existence of the high-energy electronic transition, according to  second-order perturbation theory, as previously shown by Shao and coworkers, \cite{Yang2021QEDFT} the polariton frequencies are modified to
   \begin{equation}\label{eq:omega_pm_shift}
   	\omega_{\pm}' = \omega_{\pm} - \red{\frac{g^{2}_{\rm e}}{2(\omega_{\rm e} - \omega_{\pm})} }
   \end{equation}
   Because $\omega_{\rm e} \gg \omega_{\pm}$, the high-energy electronic transition would redshift the vibrational polaritons, as observed in Fig. \ref{fig:mun_incav}b compared to Fig. \ref{fig:mun_incav}d. The high-energy electronic transitions predicted in semiclassical RT-NEO-TDDFT probably arise from approximations underlying the light-matter Hamiltonian and are not physically meaningful. In contrast, if the electronic BO approximation is applied, because the electrons are always quenched to the ground state, the influence of the unphysical high-energy electronic excitations is eliminated, thus preserving the symmetry of vibrational polaritons. \red{Interestingly, previous work  suggests that the adiabatic approximation in TDDFT may also cause spurious frequency shifts when the electronic density is significantly perturbed from the ground state. \cite{Neepa2016JCP}} 
   
   \red{Beyond the above qualitative analysis, Eq. \eqref{eq:omega_pm_shift} can also be used to estimate the magnitude of the unphysical redshift of polariton frequencies without the electronic BO approximation. Since the vibrational Rabi splitting is $\sim 100$ cm$^{-1}$, $g_{\rm v}\sim 50$ cm$^{-1}$. Because the electronic transition dipole moment is usually much larger than the vibrational transition dipole moment, we can estimate $g_{\rm e} \sim 500$ cm$^{-1}$. By further estimating $\omega_{\rm e} \sim 10^{4}$ cm$^{-1}$, we find $g_{\rm e}^2/2(\omega_{\rm e} - \omega_{\pm}) \sim 10$ cm$^{-1}$, in agreement with the frequency shift of the LP. This analysis supports our explanation that the unphysical asymmetric Rabi splitting is due to the highly excited electronic states.}
   Note that the single-molecule vibrational strong coupling example described by Fig. \ref{fig:mun_incav} is used to test our methods and should not be used to interpret polariton experiments in the collective regime.

   \subsection{RT-NEO-Ehrenfest vs BO-RT-NEO-Ehrenfest dynamics}

   \begin{figure}
   	\centering
   	\includegraphics[width=1.0\columnwidth]{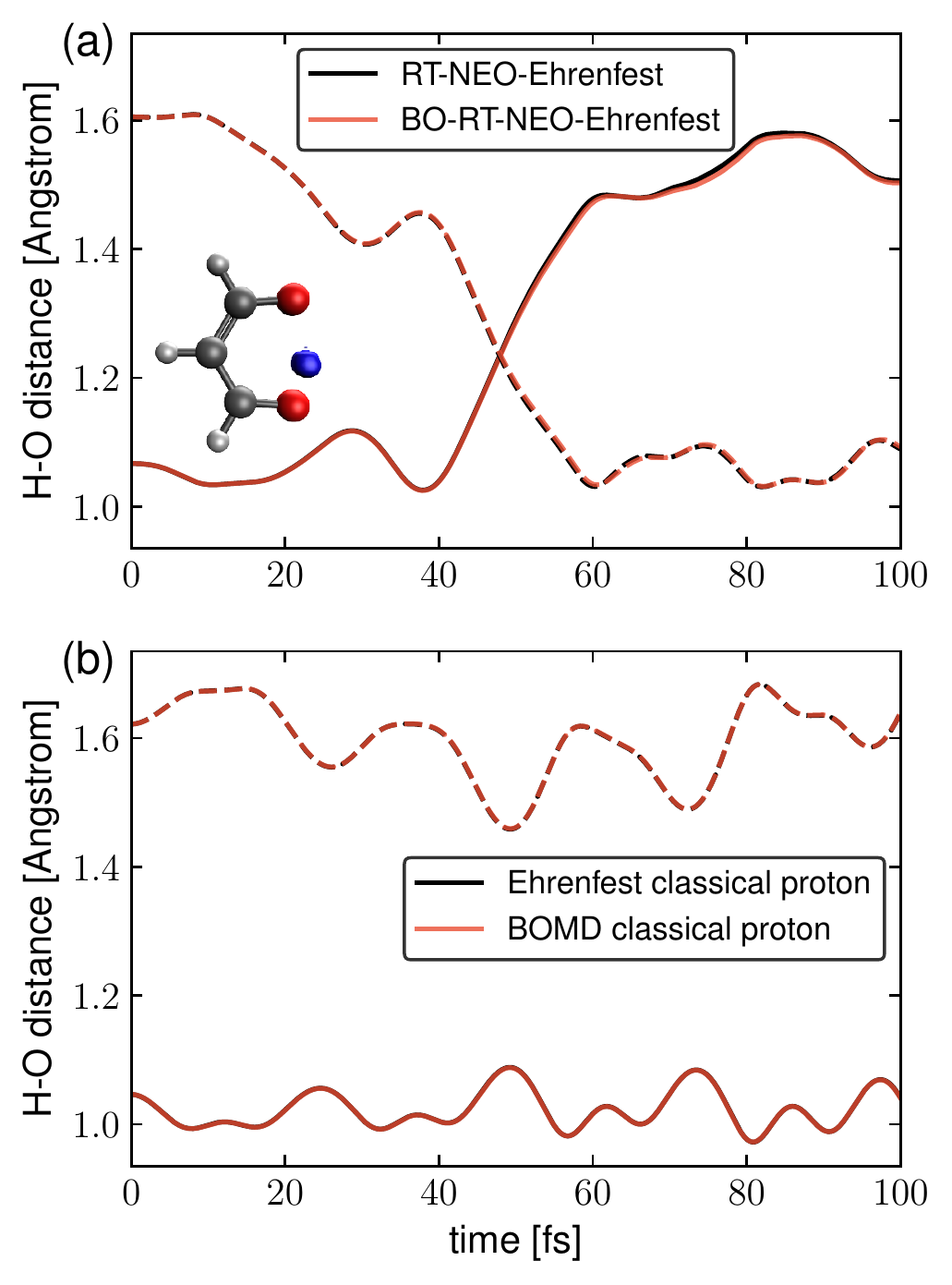}
   	\caption{(a) Electronic ground state proton transfer dynamics in malonaldehyde. The initial geometry and proton density (in blue) are shown in the inset. Two approaches are compared: (i) RT-NEO-Ehrenfest dynamics (black, with $\Delta t_{\rm q}$ = 0.010 fs) and (ii) BO-RT-NEO-Ehrenfest dynamics (red, with $\Delta t_{\rm q}$ = 0.102 fs). The quantum proton is described by three fixed proton basis function centers spanning the region between O$_{\rm D}$ and O$_{\rm A}$. The solid (dashed) lines denote the distance between the donor and acceptor oxygen atoms O$_{\rm D}$ (O$_{\rm A}$) and the expectation value of the quantum proton position. (b) The corresponding plot shown in (a) for the case when the quantum proton is replaced by a classical proton. In this case, both Ehrenfest dynamics (black) and BO molecular dynamics (red) predict no proton transfer, emphasizing the importance of a quantum treatment of the transferring proton. In both plots, the black and red lines are virtually indistinguishable.
   	}
   	\label{fig:proton_transfer}
   \end{figure}
   
   Intramolecular proton transfer in  malonaldehyde has been extensively studied both experimentally \cite{Baughcum1984,Baba1999} and theoretically. \cite{Barone1996,Tuckerman2001,Tautermann2002}
   Within the NEO framework, the transferring proton is treated quantum mechanically, and the remaining nuclei are treated classically. For our simulation, the classical nuclear geometry is chosen to be symmetric, obtained by averaging the equilibrium reactant and product geometries. The quantum proton is described by three proton basis function centers spanning the region between the donor and acceptor oxygen atoms. As shown in the inset of Fig. \ref{fig:proton_transfer}a, the quantum proton density is chosen to be the NEO-DFT SCF solution localized near the donor oxygen, O$_{\rm D}$. The proton remains localized at the NEO-DFT level because a multireference treatment is required to produce a delocalized, bilobal proton density.\cite{Yu2022MSDFT} Note that this specific initial nuclear geometry and proton density serve as an illustration of the method and are not relevant to experimental studies of this molecule. 
   
   As shown in Fig. \ref{fig:proton_transfer}a, RT-NEO-Ehrenfest dynamics (black lines) predicts proton transfer from O$_{\rm D}$ to O$_{\rm A}$ within $t = 100$ fs. This fast proton transfer reaction is induced by the nonequilibrium initial geometry. Specifically, the initial heavy nuclear geometry is symmetric, whereas the equilibrium geometry is asymmetric, with the proton bonded to one of the oxygen atoms. Hence, the classical nuclei experience forces toward the asymmetric equilibrium geometry, accompanied by nonequilibrium quantum dynamics of the transferring proton. If the proton transfer time is defined as the time when the \ch{H-O_{\rm D}} distance (solid lines) is the same as the \ch{H-O_{\rm A}} distance (dashed lines), RT-NEO-Ehrenfest dynamics (black lines) predicts that the proton transfer occurs at $t = 48$ fs.\footnote{Note that a more general definition of proton transfer time would require the proton to form a stable bond with the acceptor oxygen.} The BO-RT-NEO-Ehrenfest dynamics (red lines) is virtually identical to the RT-NEO-Ehrenfest dynamics, although the time step for BO-RT-NEO-Ehrenfest ($\Delta t_{\rm q}$ = 0.102 fs) is 10-folder larger than that for RT-NEO-Ehrenfest dynamics ($\Delta t_{\rm q}$ = 0.010 fs). 
   
\begin{figure*}
   	\centering
   	\includegraphics[width=1.0\linewidth]{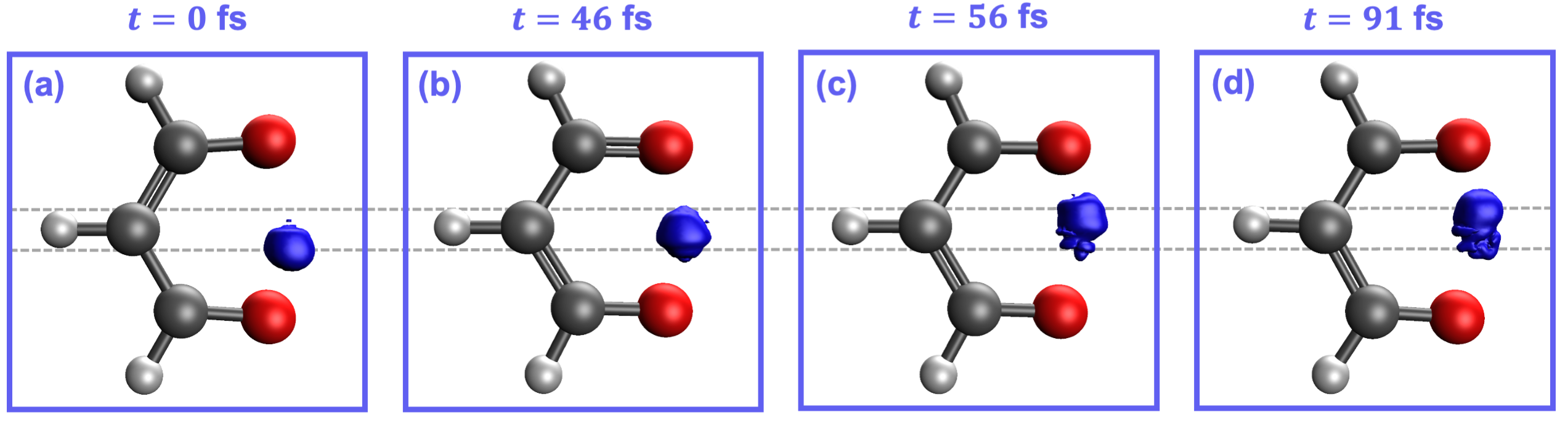}
   	\caption{Quantum proton density dynamics predicted by the BO-RT-NEO-Ehrenfest approach for the trajectory shown in Fig. \ref{fig:proton_transfer}a. Virtually identical proton densities are predicted by the RT-NEO-Ehrenfest approach. The dashed horizontal gray lines indicate the approximate equilibrium proton positions near O$_{\rm D}$ and O$_{\rm A}$. During and after proton transfer, the delocalization of proton density between O$_{\rm D}$ and O$_{\rm A}$ is captured. The quantum proton isosurface is plotted by setting the isovalue as 0.001 a.u. Note that the depiction of double bonds between classical nuclei at different times is due to the automatic rendering of the \texttt{IQMol} software and is not meaningful.
   	}
   	\label{fig:proton_transfer_cartoon}
   \end{figure*}

   For a better understanding of the proton transfer dynamics, Fig. \ref{fig:proton_transfer_cartoon} further depicts the nonequilibrium proton density at different times along the trajectory shown in Fig. \ref{fig:proton_transfer}a. Interestingly, during and after proton transfer, the proton can become delocalized in the region between \ch{O_{\rm D}} and \ch{O_{\rm A}}. This delocalization is enabled by the use of three proton basis function centers spanning the region that is sampled. A more accurate description of hydrogen tunneling is expected to require a multireference NEO approach, such as multistate DFT (NEO-MSDFT) \cite{Yu2022MSDFT} or a complete active space self-consistent-field (NEO-CASSCF) method. \cite{Webb2002}
   
   If the quantum proton is replaced by a classical proton and all the other conditions are the same as in Fig. \ref{fig:proton_transfer}a, both Ehrenfest dynamics (black lines) and BO molecular dynamics (red lines) predict no proton transfer (Fig. \ref{fig:proton_transfer}b). The difference between Fig. \ref{fig:proton_transfer}a and Fig. \ref{fig:proton_transfer}b highlights the importance of a quantum mechanical treatment of the transferring proton for proton transfer reactions, especially when hydrogen tunneling is significant. Note that the quantization of other modes for malonaldehyde has been shown to be important for a quantitatively accurate description of hydrogen tunneling \cite{Tuckerman2001} and will be investigated within the NEO framework in future studies. The total energy along the trajectories shown in  Fig. \ref{fig:proton_transfer} is plotted  in the Supplementary Material. 

\section{Conclusion}

In this manuscript, we have explored the electronic BO approximation in three different flavors of NEO dynamics: (i) RT-NEO for proton dynamics with fixed classical nuclei, (ii) semiclassical RT-NEO for polariton dynamics with fixed classical nuclei, and (iii) RT-NEO-Ehrenfest dynamics for  full molecular dynamics of molecular systems. When the BO approximation between the electrons and quantum protons is not invoked, the electronic and protonic dynamics are propagated on the same footing. Because the electronic dynamics is much faster than the protonic dynamics, a small time step is needed for a converged result. By invoking the electronic BO approximation, which involves quenching the electronic density to the SCF ground state at each time step, we can use an order-of-magnitude larger time step to perform the calculations, thus greatly reducing the computational cost. We emphasize that with this treatment, because the proton density is still propagated in real time, the nonequilibrium quantum dynamics of the proton is preserved. 
Moreover, we have also found that under vibrational strong coupling, the unphysical asymmetric Rabi splitting observed in previous semiclassical RT-NEO simulations of polaritonic systems\cite{Li2022JCTCNEO} can be fixed by invoking the electronic BO approximation, demonstrating another significant advantage of this treatment.

The application to intramolecular proton transfer in malonaldehyde highlights the importance of treating the transferring proton quantum mechanically. In this simulation,  a quantum  treatment leads to proton transfer, whereas a classical treatment predicts no proton transfer. The ability to capture proton delocalization in this proton transfer system also demonstrates the capacity of the (BO)-RT-NEO-Ehrenfest approach for more exciting applications. We emphasize that the BO-RT-NEO methods are only applicable for electronically adiabatic systems.  Thus, this approximation should not be used when there are significant non-BO effects between the electrons and quantum nuclei, as in some proton-coupled electron transfer reactions.\cite{SHS2015PCET} \red{Beyond quantum proton dynamics on the instantaneous electronic ground state, quantum proton dynamics can also be evolved on an instantaneous adiabatic electronic excited state for simulating photoexcited chemistry, such as excited state proton transfer.} Compared with the widely used ring polymer molecular dynamics approach \cite{Markland2018} for adiabatic nuclear quantum dynamics, the BO-RT-NEO methods provide a complementary perspective for describing vibrationally excited dynamics  and nuclear quantum coherence, which will be topics of future studies. Overall, this work lays the groundwork 
for applying RT-NEO methods to a wide range of electronically adiabatic and nonadiabatic chemical and biological systems.

\red{
\section{Supplementary Material}
See supplementary material for the initial molecular geometries, additional figures, linear-response TDDFT calculations of HCN, the detailed algorithms, and some implementation strategies. 
}

\section{ACKNOWLEDGMENTS}
This material is based upon work supported by the Air Force Office of Scientific Research under AFOSR Award No. FA9550-18-1-0134 for the polariton simulations and by the National Science Foundation Grant No. CHE-1954348 for the general NEO method developments. We thank
Jonathan Fetherolf, Chris Malbon,  Mathew Chow, Joseph Dickinson, and Eno Paenurk for useful discussions.

\section{Data Availability Statement}
The data and plotting scripts that support the findings of this study \red{are openly accessible at Github: \url{https://github.com/TaoELi/semiclassical-rt-neo}.}


\begin{thebibliography}{62}%
		\makeatletter
		\providecommand \@ifxundefined [1]{%
			\@ifx{#1\undefined}
		}%
		\providecommand \@ifnum [1]{%
			\ifnum #1\expandafter \@firstoftwo
			\else \expandafter \@secondoftwo
			\fi
		}%
		\providecommand \@ifx [1]{%
			\ifx #1\expandafter \@firstoftwo
			\else \expandafter \@secondoftwo
			\fi
		}%
		\providecommand \natexlab [1]{#1}%
		\providecommand \enquote  [1]{``#1''}%
		\providecommand \bibnamefont  [1]{#1}%
		\providecommand \bibfnamefont [1]{#1}%
		\providecommand \citenamefont [1]{#1}%
		\providecommand \href@noop [0]{\@secondoftwo}%
		\providecommand \href [0]{\begingroup \@sanitize@url \@href}%
		\providecommand \@href[1]{\@@startlink{#1}\@@href}%
		\providecommand \@@href[1]{\endgroup#1\@@endlink}%
		\providecommand \@sanitize@url [0]{\catcode `\\12\catcode `\$12\catcode
			`\&12\catcode `\#12\catcode `\^12\catcode `\_12\catcode `\%12\relax}%
		\providecommand \@@startlink[1]{}%
		\providecommand \@@endlink[0]{}%
		\providecommand \url  [0]{\begingroup\@sanitize@url \@url }%
		\providecommand \@url [1]{\endgroup\@href {#1}{\urlprefix }}%
		\providecommand \urlprefix  [0]{URL }%
		\providecommand \Eprint [0]{\href }%
		\providecommand \doibase [0]{https://doi.org/}%
		\providecommand \selectlanguage [0]{\@gobble}%
		\providecommand \bibinfo  [0]{\@secondoftwo}%
		\providecommand \bibfield  [0]{\@secondoftwo}%
		\providecommand \translation [1]{[#1]}%
		\providecommand \BibitemOpen [0]{}%
		\providecommand \bibitemStop [0]{}%
		\providecommand \bibitemNoStop [0]{.\EOS\space}%
		\providecommand \EOS [0]{\spacefactor3000\relax}%
		\providecommand \BibitemShut  [1]{\csname bibitem#1\endcsname}%
		\let\auto@bib@innerbib\@empty
		\bibitem [{\citenamefont {Miller}(2001)}]{Miller2001}%
		\BibitemOpen
		\bibfield  {author} {\bibinfo {author} {\bibfnamefont {W.~H.}\ \bibnamefont
				{Miller}},\ }\bibfield  {title} {\enquote {\bibinfo {title} {{The
						Semiclassical Initial Value Representation: A Potentially Practical Way for
						Adding Quantum Effects to Classical Molecular Dynamics Simulations}},}\
		}\href {https://doi.org/10.1021/jp003712k} {\bibfield  {journal} {\bibinfo
				{journal} {J. Phys. Chem. A}\ }\textbf {\bibinfo {volume} {105}},\ \bibinfo
			{pages} {2942--2955} (\bibinfo {year} {2001})}\BibitemShut {NoStop}%
		\bibitem [{\citenamefont {Cotton}\ and\ \citenamefont
			{Miller}(2013)}]{Cotton2013}%
		\BibitemOpen
		\bibfield  {author} {\bibinfo {author} {\bibfnamefont {S.~J.}\ \bibnamefont
				{Cotton}}\ and\ \bibinfo {author} {\bibfnamefont {W.~H.}\ \bibnamefont
				{Miller}},\ }\bibfield  {title} {\enquote {\bibinfo {title} {{Symmetrical
						Windowing for Quantum States in Quasi-Classical Trajectory Simulations}},}\
		}\href {https://doi.org/10.1021/jp401078u} {\bibfield  {journal} {\bibinfo
				{journal} {J. Phys. Chem. A}\ }\textbf {\bibinfo {volume} {117}},\ \bibinfo
			{pages} {7190--7194} (\bibinfo {year} {2013})}\BibitemShut {NoStop}%
		\bibitem [{\citenamefont {Markland}\ and\ \citenamefont
			{Ceriotti}(2018)}]{Markland2018}%
		\BibitemOpen
		\bibfield  {author} {\bibinfo {author} {\bibfnamefont {T.~E.}\ \bibnamefont
				{Markland}}\ and\ \bibinfo {author} {\bibfnamefont {M.}~\bibnamefont
				{Ceriotti}},\ }\bibfield  {title} {\enquote {\bibinfo {title} {{Nuclear
						quantum effects enter the mainstream}},}\ }\href
		{https://doi.org/10.1038/s41570-017-0109} {\bibfield  {journal} {\bibinfo
				{journal} {Nat. Rev. Chem.}\ }\textbf {\bibinfo {volume} {2}},\ \bibinfo
			{pages} {0109} (\bibinfo {year} {2018})}\BibitemShut {NoStop}%
		\bibitem [{\citenamefont {Worth}\ \emph {et~al.}(2008)\citenamefont {Worth},
			\citenamefont {Meyer}, \citenamefont {K{\"{o}}ppel}, \citenamefont
			{Cederbaum},\ and\ \citenamefont {Burghardt}}]{Worth2008MCTDH}%
		\BibitemOpen
		\bibfield  {author} {\bibinfo {author} {\bibfnamefont {G.~A.}\ \bibnamefont
				{Worth}}, \bibinfo {author} {\bibfnamefont {H.-D.}\ \bibnamefont {Meyer}},
			\bibinfo {author} {\bibfnamefont {H.}~\bibnamefont {K{\"{o}}ppel}}, \bibinfo
			{author} {\bibfnamefont {L.~S.}\ \bibnamefont {Cederbaum}},\ and\ \bibinfo
			{author} {\bibfnamefont {I.}~\bibnamefont {Burghardt}},\ }\bibfield  {title}
		{\enquote {\bibinfo {title} {{Using the MCTDH wavepacket propagation method
						to describe multimode non-adiabatic dynamics}},}\ }\href
		{https://doi.org/10.1080/01442350802137656} {\bibfield  {journal} {\bibinfo
				{journal} {Int. Rev. Phys. Chem.}\ }\textbf {\bibinfo {volume} {27}},\
			\bibinfo {pages} {569--606} (\bibinfo {year} {2008})}\BibitemShut {NoStop}%
		\bibitem [{\citenamefont {Abedi}, \citenamefont {Maitra},\ and\ \citenamefont
			{Gross}(2010)}]{Abedi2010EF}%
		\BibitemOpen
		\bibfield  {author} {\bibinfo {author} {\bibfnamefont {A.}~\bibnamefont
				{Abedi}}, \bibinfo {author} {\bibfnamefont {N.~T.}\ \bibnamefont {Maitra}},\
			and\ \bibinfo {author} {\bibfnamefont {E.~K.~U.}\ \bibnamefont {Gross}},\
		}\bibfield  {title} {\enquote {\bibinfo {title} {{Exact Factorization of the
						Time-Dependent Electron-Nuclear Wave Function}},}\ }\href
		{https://doi.org/10.1103/PhysRevLett.105.123002} {\bibfield  {journal}
			{\bibinfo  {journal} {Phys. Rev. Lett.}\ }\textbf {\bibinfo {volume} {105}},\
			\bibinfo {pages} {123002} (\bibinfo {year} {2010})}\BibitemShut {NoStop}%
		\bibitem [{\citenamefont {Webb}, \citenamefont {Iordanov},\ and\ \citenamefont
			{Hammes-Schiffer}(2002)}]{Webb2002}%
		\BibitemOpen
		\bibfield  {author} {\bibinfo {author} {\bibfnamefont {S.~P.}\ \bibnamefont
				{Webb}}, \bibinfo {author} {\bibfnamefont {T.}~\bibnamefont {Iordanov}},\
			and\ \bibinfo {author} {\bibfnamefont {S.}~\bibnamefont {Hammes-Schiffer}},\
		}\bibfield  {title} {\enquote {\bibinfo {title} {{Multiconfigurational
						nuclear-electronic orbital approach: Incorporation of nuclear quantum effects
						in electronic structure calculations}},}\ }\href
		{https://doi.org/10.1063/1.1494980} {\bibfield  {journal} {\bibinfo
				{journal} {J. Chem. Phys.}\ }\textbf {\bibinfo {volume} {117}},\ \bibinfo
			{pages} {4106--4118} (\bibinfo {year} {2002})}\BibitemShut {NoStop}%
		\bibitem [{\citenamefont {Pavo{\v{s}}evi{\'{c}}}, \citenamefont {Culpitt},\
			and\ \citenamefont {Hammes-Schiffer}(2020)}]{Pavosevic2020}%
		\BibitemOpen
		\bibfield  {author} {\bibinfo {author} {\bibfnamefont {F.}~\bibnamefont
				{Pavo{\v{s}}evi{\'{c}}}}, \bibinfo {author} {\bibfnamefont {T.}~\bibnamefont
				{Culpitt}},\ and\ \bibinfo {author} {\bibfnamefont {S.}~\bibnamefont
				{Hammes-Schiffer}},\ }\bibfield  {title} {\enquote {\bibinfo {title}
				{{Multicomponent Quantum Chemistry: Integrating Electronic and Nuclear
						Quantum Effects via the Nuclear–Electronic Orbital Method}},}\ }\href
		{https://doi.org/10.1021/acs.chemrev.9b00798} {\bibfield  {journal} {\bibinfo
				{journal} {Chem. Rev.}\ }\textbf {\bibinfo {volume} {120}},\ \bibinfo
			{pages} {4222--4253} (\bibinfo {year} {2020})}\BibitemShut {NoStop}%
		\bibitem [{\citenamefont {Tao}\ \emph {et~al.}(2021)\citenamefont {Tao},
			\citenamefont {Yu}, \citenamefont {Roy},\ and\ \citenamefont
			{Hammes-Schiffer}}]{Tao2021}%
		\BibitemOpen
		\bibfield  {author} {\bibinfo {author} {\bibfnamefont {Z.}~\bibnamefont
				{Tao}}, \bibinfo {author} {\bibfnamefont {Q.}~\bibnamefont {Yu}}, \bibinfo
			{author} {\bibfnamefont {S.}~\bibnamefont {Roy}},\ and\ \bibinfo {author}
			{\bibfnamefont {S.}~\bibnamefont {Hammes-Schiffer}},\ }\bibfield  {title}
		{\enquote {\bibinfo {title} {{Direct Dynamics with Nuclear–Electronic
						Orbital Density Functional Theory}},}\ }\href
		{https://doi.org/10.1021/acs.accounts.1c00516} {\bibfield  {journal}
			{\bibinfo  {journal} {Acc. Chem. Res.}\ }\textbf {\bibinfo {volume} {54}},\
			\bibinfo {pages} {4131--4141} (\bibinfo {year} {2021})}\BibitemShut {NoStop}%
		\bibitem [{\citenamefont {Xu}, \citenamefont {Chen},\ and\ \citenamefont
			{Yang}(2022)}]{Xu2022}%
		\BibitemOpen
		\bibfield  {author} {\bibinfo {author} {\bibfnamefont {X.}~\bibnamefont
				{Xu}}, \bibinfo {author} {\bibfnamefont {Z.}~\bibnamefont {Chen}},\ and\
			\bibinfo {author} {\bibfnamefont {Y.}~\bibnamefont {Yang}},\ }\bibfield
		{title} {\enquote {\bibinfo {title} {{Molecular Dynamics with Constrained
						Nuclear Electronic Orbital Density Functional Theory: Accurate Vibrational
						Spectra from Efficient Incorporation of Nuclear Quantum Effects}},}\ }\href
		{https://doi.org/10.1021/jacs.1c12932} {\bibfield  {journal} {\bibinfo
				{journal} {J. Am. Chem. Soc.}\ }\textbf {\bibinfo {volume} {144}},\ \bibinfo
			{pages} {4039--4046} (\bibinfo {year} {2022})}\BibitemShut {NoStop}%
		\bibitem [{\citenamefont {Yu}, \citenamefont {Roy},\ and\ \citenamefont
			{Hammes-Schiffer}(2022)}]{Yu2022MSDFT}%
		\BibitemOpen
		\bibfield  {author} {\bibinfo {author} {\bibfnamefont {Q.}~\bibnamefont
				{Yu}}, \bibinfo {author} {\bibfnamefont {S.}~\bibnamefont {Roy}},\ and\
			\bibinfo {author} {\bibfnamefont {S.}~\bibnamefont {Hammes-Schiffer}},\
		}\bibfield  {title} {\enquote {\bibinfo {title} {{Nonadiabatic Dynamics of
						Hydrogen Tunneling with Nuclear-Electronic Orbital Multistate Density
						Functional Theory}},}\ }\href {https://doi.org/10.1021/acs.jctc.2c00938}
		{\bibfield  {journal} {\bibinfo  {journal} {J. Chem. Theory Comput.}\ }
			(\bibinfo {year} {2022}),\ 10.1021/acs.jctc.2c00938}\BibitemShut {NoStop}%
		\bibitem [{\citenamefont {Zhao}\ \emph
			{et~al.}(2020{\natexlab{a}})\citenamefont {Zhao}, \citenamefont {Tao},
			\citenamefont {Pavo{\v{s}}evi{\'{c}}}, \citenamefont {Wildman}, \citenamefont
			{Hammes-Schiffer},\ and\ \citenamefont {Li}}]{Zhao2020}%
		\BibitemOpen
		\bibfield  {author} {\bibinfo {author} {\bibfnamefont {L.}~\bibnamefont
				{Zhao}}, \bibinfo {author} {\bibfnamefont {Z.}~\bibnamefont {Tao}}, \bibinfo
			{author} {\bibfnamefont {F.}~\bibnamefont {Pavo{\v{s}}evi{\'{c}}}}, \bibinfo
			{author} {\bibfnamefont {A.}~\bibnamefont {Wildman}}, \bibinfo {author}
			{\bibfnamefont {S.}~\bibnamefont {Hammes-Schiffer}},\ and\ \bibinfo {author}
			{\bibfnamefont {X.}~\bibnamefont {Li}},\ }\bibfield  {title} {\enquote
			{\bibinfo {title} {{Real-Time Time-Dependent Nuclear-Electronic Orbital
						Approach: Dynamics beyond the Born-Oppenheimer Approximation}},}\ }\href
		{https://doi.org/10.1021/ACS.JPCLETT.0C00701/SUPPL_FILE/JZ0C00701_SI_002.ZIP}
		{\bibfield  {journal} {\bibinfo  {journal} {J. Phys. Chem. Lett.}\ }\textbf
			{\bibinfo {volume} {11}},\ \bibinfo {pages} {4052--4058} (\bibinfo {year}
			{2020}{\natexlab{a}})}\BibitemShut {NoStop}%
		\bibitem [{\citenamefont {Zhao}\ \emph
			{et~al.}(2020{\natexlab{b}})\citenamefont {Zhao}, \citenamefont {Wildman},
			\citenamefont {Tao}, \citenamefont {Schneider}, \citenamefont
			{Hammes-Schiffer},\ and\ \citenamefont {Li}}]{Zhao2020JCP}%
		\BibitemOpen
		\bibfield  {author} {\bibinfo {author} {\bibfnamefont {L.}~\bibnamefont
				{Zhao}}, \bibinfo {author} {\bibfnamefont {A.}~\bibnamefont {Wildman}},
			\bibinfo {author} {\bibfnamefont {Z.}~\bibnamefont {Tao}}, \bibinfo {author}
			{\bibfnamefont {P.}~\bibnamefont {Schneider}}, \bibinfo {author}
			{\bibfnamefont {S.}~\bibnamefont {Hammes-Schiffer}},\ and\ \bibinfo {author}
			{\bibfnamefont {X.}~\bibnamefont {Li}},\ }\bibfield  {title} {\enquote
			{\bibinfo {title} {{Nuclear–electronic orbital Ehrenfest dynamics}},}\
		}\href {https://doi.org/10.1063/5.0031019} {\bibfield  {journal} {\bibinfo
				{journal} {J. Chem. Phys.}\ }\textbf {\bibinfo {volume} {153}},\ \bibinfo
			{pages} {224111} (\bibinfo {year} {2020}{\natexlab{b}})}\BibitemShut
		{NoStop}%
		\bibitem [{\citenamefont {Zhao}\ \emph {et~al.}(2021)\citenamefont {Zhao},
			\citenamefont {Wildman}, \citenamefont {Pavo{\v{s}}evi{\'{c}}}, \citenamefont
			{Tully}, \citenamefont {Hammes-Schiffer},\ and\ \citenamefont
			{Li}}]{Zhao2021JPCL}%
		\BibitemOpen
		\bibfield  {author} {\bibinfo {author} {\bibfnamefont {L.}~\bibnamefont
				{Zhao}}, \bibinfo {author} {\bibfnamefont {A.}~\bibnamefont {Wildman}},
			\bibinfo {author} {\bibfnamefont {F.}~\bibnamefont {Pavo{\v{s}}evi{\'{c}}}},
			\bibinfo {author} {\bibfnamefont {J.~C.}\ \bibnamefont {Tully}}, \bibinfo
			{author} {\bibfnamefont {S.}~\bibnamefont {Hammes-Schiffer}},\ and\ \bibinfo
			{author} {\bibfnamefont {X.}~\bibnamefont {Li}},\ }\bibfield  {title}
		{\enquote {\bibinfo {title} {{Excited State Intramolecular Proton Transfer
						with Nuclear-Electronic Orbital Ehrenfest Dynamics}},}\ }\href
		{https://doi.org/10.1021/acs.jpclett.1c00564} {\bibfield  {journal} {\bibinfo
				{journal} {J. Phys. Chem. Lett.}\ }\textbf {\bibinfo {volume} {12}},\
			\bibinfo {pages} {3497--3502} (\bibinfo {year} {2021})}\BibitemShut {NoStop}%
		\bibitem [{\citenamefont {Li}, \citenamefont {Tao},\ and\ \citenamefont
			{Hammes-Schiffer}(2022)}]{Li2022JCTCNEO}%
		\BibitemOpen
		\bibfield  {author} {\bibinfo {author} {\bibfnamefont {T.~E.}\ \bibnamefont
				{Li}}, \bibinfo {author} {\bibfnamefont {Z.}~\bibnamefont {Tao}},\ and\
			\bibinfo {author} {\bibfnamefont {S.}~\bibnamefont {Hammes-Schiffer}},\
		}\bibfield  {title} {\enquote {\bibinfo {title} {{Semiclassical Real-Time
						Nuclear-Electronic Orbital Dynamics for Molecular Polaritons: Unified Theory
						of Electronic and Vibrational Strong Couplings}},}\ }\href
		{https://doi.org/10.1021/acs.jctc.2c00096} {\bibfield  {journal} {\bibinfo
				{journal} {J. Chem. Theory Comput.}\ }\textbf {\bibinfo {volume} {18}},\
			\bibinfo {pages} {2774--2784} (\bibinfo {year} {2022})},\ \Eprint
		{https://arxiv.org/abs/2203.04952} {arXiv:2203.04952} \BibitemShut {NoStop}%
		\bibitem [{\citenamefont {Li}\ \emph {et~al.}(2005{\natexlab{a}})\citenamefont
			{Li}, \citenamefont {Tully}, \citenamefont {Schlegel},\ and\ \citenamefont
			{Frisch}}]{Li2005Eh}%
		\BibitemOpen
		\bibfield  {author} {\bibinfo {author} {\bibfnamefont {X.}~\bibnamefont
				{Li}}, \bibinfo {author} {\bibfnamefont {J.~C.}\ \bibnamefont {Tully}},
			\bibinfo {author} {\bibfnamefont {H.~B.}\ \bibnamefont {Schlegel}},\ and\
			\bibinfo {author} {\bibfnamefont {M.~J.}\ \bibnamefont {Frisch}},\ }\bibfield
		{title} {\enquote {\bibinfo {title} {{Ab initio Ehrenfest dynamics}},}\
		}\href {https://doi.org/10.1063/1.2008258} {\bibfield  {journal} {\bibinfo
				{journal} {J. Chem. Phys.}\ }\textbf {\bibinfo {volume} {123}},\ \bibinfo
			{pages} {084106} (\bibinfo {year} {2005}{\natexlab{a}})}\BibitemShut
		{NoStop}%
		\bibitem [{\citenamefont {Isborn}, \citenamefont {Li},\ and\ \citenamefont
			{Tully}(2007)}]{Isborn2007}%
		\BibitemOpen
		\bibfield  {author} {\bibinfo {author} {\bibfnamefont {C.~M.}\ \bibnamefont
				{Isborn}}, \bibinfo {author} {\bibfnamefont {X.}~\bibnamefont {Li}},\ and\
			\bibinfo {author} {\bibfnamefont {J.~C.}\ \bibnamefont {Tully}},\ }\bibfield
		{title} {\enquote {\bibinfo {title} {{Time-dependent density functional
						theory Ehrenfest dynamics: Collisions between atomic oxygen and graphite
						clusters}},}\ }\href {https://doi.org/10.1063/1.2713391} {\bibfield
			{journal} {\bibinfo  {journal} {J. Chem. Phys.}\ }\textbf {\bibinfo {volume}
				{126}},\ \bibinfo {pages} {134307} (\bibinfo {year} {2007})}\BibitemShut
		{NoStop}%
		\bibitem [{\citenamefont {Tully}(1990)}]{Tully1990}%
		\BibitemOpen
		\bibfield  {author} {\bibinfo {author} {\bibfnamefont {J.~C.}\ \bibnamefont
				{Tully}},\ }\bibfield  {title} {\enquote {\bibinfo {title} {{Molecular
						Dynamics with Electronic Transitions}},}\ }\href
		{https://doi.org/10.1063/1.459170} {\bibfield  {journal} {\bibinfo  {journal}
				{J. Chem. Phys.}\ }\textbf {\bibinfo {volume} {93}},\ \bibinfo {pages}
			{1061--1071} (\bibinfo {year} {1990})}\BibitemShut {NoStop}%
		\bibitem [{\citenamefont {Li}\ and\ \citenamefont
			{Tong}(1986)}]{Li1986TDDFTMulti}%
		\BibitemOpen
		\bibfield  {author} {\bibinfo {author} {\bibfnamefont {T.-C.}\ \bibnamefont
				{Li}}\ and\ \bibinfo {author} {\bibfnamefont {P.-Q.}\ \bibnamefont {Tong}},\
		}\bibfield  {title} {\enquote {\bibinfo {title} {Time-dependent
					density-functional theory for multicomponent systems},}\ }\href
		{https://doi.org/10.1103/physreva.34.529} {\bibfield  {journal} {\bibinfo
				{journal} {Phys. Rev. A}\ }\textbf {\bibinfo {volume} {34}},\ \bibinfo
			{pages} {529--532} (\bibinfo {year} {1986})}\BibitemShut {NoStop}%
		\bibitem [{\citenamefont {van Leeuwen}\ and\ \citenamefont
			{Gross}(2006)}]{Marques2006TDDFT}%
		\BibitemOpen
		\bibfield  {author} {\bibinfo {author} {\bibfnamefont {R.}~\bibnamefont {van
					Leeuwen}}\ and\ \bibinfo {author} {\bibfnamefont {E.~K.~U.}\ \bibnamefont
				{Gross}},\ }\bibfield  {title} {\enquote {\bibinfo {title} {Multicomponent
					density-functional theory},}\ }in\ \href {https://doi.org/10.1007/b11767107}
		{\emph {\bibinfo {booktitle} {Time-Dependent Density Functional Theory}}},\
		\bibinfo {editor} {edited by\ \bibinfo {editor} {\bibfnamefont {M.~A.}\
				\bibnamefont {Marques}}, \bibinfo {editor} {\bibfnamefont {C.~A.}\
				\bibnamefont {Ullrich}}, \bibinfo {editor} {\bibfnamefont {F.}~\bibnamefont
				{Nogueira}}, \bibinfo {editor} {\bibfnamefont {A.}~\bibnamefont {Rubio}},
			\bibinfo {editor} {\bibfnamefont {K.}~\bibnamefont {Burke}},\ and\ \bibinfo
			{editor} {\bibfnamefont {E.~K.~U.}\ \bibnamefont {Gross}}}\ (\bibinfo
		{publisher} {Springer Berlin Heidelberg},\ \bibinfo {year} {2006})\ pp.\
		\bibinfo {pages} {93--106}\BibitemShut {NoStop}%
		\bibitem [{\citenamefont {Butriy}\ \emph {et~al.}(2007)\citenamefont {Butriy},
			\citenamefont {Ebadi}, \citenamefont {de~Boeij}, \citenamefont {van
				Leeuwen},\ and\ \citenamefont {Gross}}]{Butriy2007}%
		\BibitemOpen
		\bibfield  {author} {\bibinfo {author} {\bibfnamefont {O.}~\bibnamefont
				{Butriy}}, \bibinfo {author} {\bibfnamefont {H.}~\bibnamefont {Ebadi}},
			\bibinfo {author} {\bibfnamefont {P.~L.}\ \bibnamefont {de~Boeij}}, \bibinfo
			{author} {\bibfnamefont {R.}~\bibnamefont {van Leeuwen}},\ and\ \bibinfo
			{author} {\bibfnamefont {E.~K.~U.}\ \bibnamefont {Gross}},\ }\bibfield
		{title} {\enquote {\bibinfo {title} {Multicomponent density-functional theory
					for time-dependent systems},}\ }\href
		{https://doi.org/10.1103/PhysRevA.76.052514} {\bibfield  {journal} {\bibinfo
				{journal} {Phys. Rev. A}\ }\textbf {\bibinfo {volume} {76}},\ \bibinfo
			{pages} {052514} (\bibinfo {year} {2007})}\BibitemShut {NoStop}%
		\bibitem [{\citenamefont {Yang}, \citenamefont {Culpitt},\ and\ \citenamefont
			{Hammes-Schiffer}(2018)}]{Yang2018}%
		\BibitemOpen
		\bibfield  {author} {\bibinfo {author} {\bibfnamefont {Y.}~\bibnamefont
				{Yang}}, \bibinfo {author} {\bibfnamefont {T.}~\bibnamefont {Culpitt}},\ and\
			\bibinfo {author} {\bibfnamefont {S.}~\bibnamefont {Hammes-Schiffer}},\
		}\bibfield  {title} {\enquote {\bibinfo {title} {{Multicomponent
						Time-Dependent Density Functional Theory: Proton and Electron Excitation
						Energies}},}\ }\href
		{https://doi.org/10.1021/ACS.JPCLETT.8B00547/SUPPL_FILE/JZ8B00547_SI_001.PDF}
		{\bibfield  {journal} {\bibinfo  {journal} {J. Phys. Chem. Lett.}\ }\textbf
			{\bibinfo {volume} {9}},\ \bibinfo {pages} {1765--1770} (\bibinfo {year}
			{2018})}\BibitemShut {NoStop}%
		\bibitem [{\citenamefont {Thomas}\ \emph {et~al.}(2019)\citenamefont {Thomas},
			\citenamefont {Lethuillier-Karl}, \citenamefont {Nagarajan}, \citenamefont
			{Vergauwe}, \citenamefont {George}, \citenamefont {Chervy}, \citenamefont
			{Shalabney}, \citenamefont {Devaux}, \citenamefont {Genet}, \citenamefont
			{Moran},\ and\ \citenamefont {Ebbesen}}]{Thomas2019_science}%
		\BibitemOpen
		\bibfield  {author} {\bibinfo {author} {\bibfnamefont {A.}~\bibnamefont
				{Thomas}}, \bibinfo {author} {\bibfnamefont {L.}~\bibnamefont
				{Lethuillier-Karl}}, \bibinfo {author} {\bibfnamefont {K.}~\bibnamefont
				{Nagarajan}}, \bibinfo {author} {\bibfnamefont {R.~M.~A.}\ \bibnamefont
				{Vergauwe}}, \bibinfo {author} {\bibfnamefont {J.}~\bibnamefont {George}},
			\bibinfo {author} {\bibfnamefont {T.}~\bibnamefont {Chervy}}, \bibinfo
			{author} {\bibfnamefont {A.}~\bibnamefont {Shalabney}}, \bibinfo {author}
			{\bibfnamefont {E.}~\bibnamefont {Devaux}}, \bibinfo {author} {\bibfnamefont
				{C.}~\bibnamefont {Genet}}, \bibinfo {author} {\bibfnamefont
				{J.}~\bibnamefont {Moran}},\ and\ \bibinfo {author} {\bibfnamefont {T.~W.}\
				\bibnamefont {Ebbesen}},\ }\bibfield  {title} {\enquote {\bibinfo {title}
				{{Tilting a Ground-State Reactivity Landscape by Vibrational Strong
						Coupling}},}\ }\href {https://doi.org/10.1126/science.aau7742} {\bibfield
			{journal} {\bibinfo  {journal} {Science}\ }\textbf {\bibinfo {volume}
				{363}},\ \bibinfo {pages} {615--619} (\bibinfo {year} {2019})}\BibitemShut
		{NoStop}%
		\bibitem [{\citenamefont {Li}\ \emph {et~al.}(2022)\citenamefont {Li},
			\citenamefont {Cui}, \citenamefont {Subotnik},\ and\ \citenamefont
			{Nitzan}}]{Li2022Review}%
		\BibitemOpen
		\bibfield  {author} {\bibinfo {author} {\bibfnamefont {T.~E.}\ \bibnamefont
				{Li}}, \bibinfo {author} {\bibfnamefont {B.}~\bibnamefont {Cui}}, \bibinfo
			{author} {\bibfnamefont {J.~E.}\ \bibnamefont {Subotnik}},\ and\ \bibinfo
			{author} {\bibfnamefont {A.}~\bibnamefont {Nitzan}},\ }\bibfield  {title}
		{\enquote {\bibinfo {title} {{Molecular Polaritonics: Chemical Dynamics Under
						Strong Light–Matter Coupling}},}\ }\href
		{https://doi.org/10.1146/annurev-physchem-090519-042621} {\bibfield
			{journal} {\bibinfo  {journal} {Annu. Rev. Phys. Chem.}\ }\textbf {\bibinfo
				{volume} {73}} (\bibinfo {year} {2022}),\
			10.1146/annurev-physchem-090519-042621}\BibitemShut {NoStop}%
		\bibitem [{\citenamefont {Fregoni}, \citenamefont {Garcia-Vidal},\ and\
			\citenamefont {Feist}(2022)}]{Fregoni2022}%
		\BibitemOpen
		\bibfield  {author} {\bibinfo {author} {\bibfnamefont {J.}~\bibnamefont
				{Fregoni}}, \bibinfo {author} {\bibfnamefont {F.~J.}\ \bibnamefont
				{Garcia-Vidal}},\ and\ \bibinfo {author} {\bibfnamefont {J.}~\bibnamefont
				{Feist}},\ }\bibfield  {title} {\enquote {\bibinfo {title} {{Theoretical
						Challenges in Polaritonic Chemistry}},}\ }\href
		{https://doi.org/10.1021/acsphotonics.1c01749} {\bibfield  {journal}
			{\bibinfo  {journal} {ACS Photonics}\ }\textbf {\bibinfo {volume} {9}},\
			\bibinfo {pages} {1096--1107} (\bibinfo {year} {2022})}\BibitemShut {NoStop}%
		\bibitem [{\citenamefont {Nagarajan}, \citenamefont {Thomas},\ and\
			\citenamefont {Ebbesen}(2021)}]{Nagarajan2021}%
		\BibitemOpen
		\bibfield  {author} {\bibinfo {author} {\bibfnamefont {K.}~\bibnamefont
				{Nagarajan}}, \bibinfo {author} {\bibfnamefont {A.}~\bibnamefont {Thomas}},\
			and\ \bibinfo {author} {\bibfnamefont {T.~W.}\ \bibnamefont {Ebbesen}},\
		}\bibfield  {title} {\enquote {\bibinfo {title} {{Chemistry under Vibrational
						Strong Coupling}},}\ }\href {https://doi.org/10.1021/jacs.1c07420} {\bibfield
			{journal} {\bibinfo  {journal} {J. Am. Chem. Soc.}\ }\textbf {\bibinfo
				{volume} {143}},\ \bibinfo {pages} {16877--16889} (\bibinfo {year}
			{2021})}\BibitemShut {NoStop}%
		\bibitem [{\citenamefont {Xu}\ and\ \citenamefont {Yang}(2020)}]{Xu2020}%
		\BibitemOpen
		\bibfield  {author} {\bibinfo {author} {\bibfnamefont {X.}~\bibnamefont
				{Xu}}\ and\ \bibinfo {author} {\bibfnamefont {Y.}~\bibnamefont {Yang}},\
		}\bibfield  {title} {\enquote {\bibinfo {title} {{Constrained
						nuclear-electronic orbital density functional theory: Energy surfaces with
						nuclear quantum effects}},}\ }\href {https://doi.org/10.1063/1.5143371}
		{\bibfield  {journal} {\bibinfo  {journal} {J. Chem. Phys.}\ }\textbf
			{\bibinfo {volume} {152}},\ \bibinfo {pages} {084107} (\bibinfo {year}
			{2020})}\BibitemShut {NoStop}%
		\bibitem [{\citenamefont {Baughcum}\ \emph {et~al.}(1984)\citenamefont
			{Baughcum}, \citenamefont {Smith}, \citenamefont {Wilson},\ and\
			\citenamefont {Duerst}}]{Baughcum1984}%
		\BibitemOpen
		\bibfield  {author} {\bibinfo {author} {\bibfnamefont {S.~L.}\ \bibnamefont
				{Baughcum}}, \bibinfo {author} {\bibfnamefont {Z.}~\bibnamefont {Smith}},
			\bibinfo {author} {\bibfnamefont {E.~B.}\ \bibnamefont {Wilson}},\ and\
			\bibinfo {author} {\bibfnamefont {R.~W.}\ \bibnamefont {Duerst}},\ }\bibfield
		{title} {\enquote {\bibinfo {title} {{Microwave spectroscopic study of
						malonaldehyde. 3. Vibration-rotation interaction and one-dimensional model
						for proton tunneling}},}\ }\href {https://doi.org/10.1021/ja00320a007}
		{\bibfield  {journal} {\bibinfo  {journal} {J. Am. Chem. Soc.}\ }\textbf
			{\bibinfo {volume} {106}},\ \bibinfo {pages} {2260--2265} (\bibinfo {year}
			{1984})}\BibitemShut {NoStop}%
		\bibitem [{\citenamefont {Baba}\ \emph {et~al.}(1999)\citenamefont {Baba},
			\citenamefont {Tanaka}, \citenamefont {Morino}, \citenamefont {Yamada},\ and\
			\citenamefont {Tanaka}}]{Baba1999}%
		\BibitemOpen
		\bibfield  {author} {\bibinfo {author} {\bibfnamefont {T.}~\bibnamefont
				{Baba}}, \bibinfo {author} {\bibfnamefont {T.}~\bibnamefont {Tanaka}},
			\bibinfo {author} {\bibfnamefont {I.}~\bibnamefont {Morino}}, \bibinfo
			{author} {\bibfnamefont {K.~M.~T.}\ \bibnamefont {Yamada}},\ and\ \bibinfo
			{author} {\bibfnamefont {K.}~\bibnamefont {Tanaka}},\ }\bibfield  {title}
		{\enquote {\bibinfo {title} {{Detection of the tunneling-rotation transitions
						of malonaldehyde in the submillimeter-wave region}},}\ }\href
		{https://doi.org/10.1063/1.478296} {\bibfield  {journal} {\bibinfo  {journal}
				{J. Chem. Phys.}\ }\textbf {\bibinfo {volume} {110}},\ \bibinfo {pages}
			{4131--4133} (\bibinfo {year} {1999})}\BibitemShut {NoStop}%
		\bibitem [{\citenamefont {Barone}\ and\ \citenamefont
			{Adamo}(1996)}]{Barone1996}%
		\BibitemOpen
		\bibfield  {author} {\bibinfo {author} {\bibfnamefont {V.}~\bibnamefont
				{Barone}}\ and\ \bibinfo {author} {\bibfnamefont {C.}~\bibnamefont {Adamo}},\
		}\bibfield  {title} {\enquote {\bibinfo {title} {{Proton transfer in the
						ground and lowest excited states of malonaldehyde: A comparative density
						functional and post‐Hartree–Fock study}},}\ }\href
		{https://doi.org/10.1063/1.472900} {\bibfield  {journal} {\bibinfo  {journal}
				{J. Chem. Phys.}\ }\textbf {\bibinfo {volume} {105}},\ \bibinfo {pages}
			{11007--11019} (\bibinfo {year} {1996})}\BibitemShut {NoStop}%
		\bibitem [{\citenamefont {Tuckerman}\ and\ \citenamefont
			{Marx}(2001)}]{Tuckerman2001}%
		\BibitemOpen
		\bibfield  {author} {\bibinfo {author} {\bibfnamefont {M.~E.}\ \bibnamefont
				{Tuckerman}}\ and\ \bibinfo {author} {\bibfnamefont {D.}~\bibnamefont
				{Marx}},\ }\bibfield  {title} {\enquote {\bibinfo {title} {{Heavy-Atom
						Skeleton Quantization and Proton Tunneling in “Intermediate-Barrier”
						Hydrogen Bonds}},}\ }\href {https://doi.org/10.1103/PhysRevLett.86.4946}
		{\bibfield  {journal} {\bibinfo  {journal} {Phys. Rev. Lett.}\ }\textbf
			{\bibinfo {volume} {86}},\ \bibinfo {pages} {4946--4949} (\bibinfo {year}
			{2001})}\BibitemShut {NoStop}%
		\bibitem [{\citenamefont {Tautermann}\ \emph {et~al.}(2002)\citenamefont
			{Tautermann}, \citenamefont {Voegele}, \citenamefont {Loerting},\ and\
			\citenamefont {Liedl}}]{Tautermann2002}%
		\BibitemOpen
		\bibfield  {author} {\bibinfo {author} {\bibfnamefont {C.~S.}\ \bibnamefont
				{Tautermann}}, \bibinfo {author} {\bibfnamefont {A.~F.}\ \bibnamefont
				{Voegele}}, \bibinfo {author} {\bibfnamefont {T.}~\bibnamefont {Loerting}},\
			and\ \bibinfo {author} {\bibfnamefont {K.~R.}\ \bibnamefont {Liedl}},\
		}\bibfield  {title} {\enquote {\bibinfo {title} {{The optimal tunneling path
						for the proton transfer in malonaldehyde}},}\ }\href
		{https://doi.org/10.1063/1.1488924} {\bibfield  {journal} {\bibinfo
				{journal} {J. Chem. Phys.}\ }\textbf {\bibinfo {volume} {117}},\ \bibinfo
			{pages} {1962--1966} (\bibinfo {year} {2002})}\BibitemShut {NoStop}%
		\bibitem [{\citenamefont {Pak}, \citenamefont {Chakraborty},\ and\
			\citenamefont {Hammes-Schiffer}(2007)}]{Pak2007}%
		\BibitemOpen
		\bibfield  {author} {\bibinfo {author} {\bibfnamefont {M.~V.}\ \bibnamefont
				{Pak}}, \bibinfo {author} {\bibfnamefont {A.}~\bibnamefont {Chakraborty}},\
			and\ \bibinfo {author} {\bibfnamefont {S.}~\bibnamefont {Hammes-Schiffer}},\
		}\bibfield  {title} {\enquote {\bibinfo {title} {{Density Functional Theory
						Treatment of Electron Correlation in the Nuclear-Electronic Orbital
						Approach}},}\ }\href {https://doi.org/10.1021/jp0704463} {\bibfield
			{journal} {\bibinfo  {journal} {J. Phys. Chem. A}\ }\textbf {\bibinfo
				{volume} {111}},\ \bibinfo {pages} {4522--4526} (\bibinfo {year}
			{2007})}\BibitemShut {NoStop}%
		\bibitem [{\citenamefont {Chakraborty}, \citenamefont {Pak},\ and\
			\citenamefont {Hammes-Schiffer}(2008)}]{Chakraborty2008}%
		\BibitemOpen
		\bibfield  {author} {\bibinfo {author} {\bibfnamefont {A.}~\bibnamefont
				{Chakraborty}}, \bibinfo {author} {\bibfnamefont {M.~V.}\ \bibnamefont
				{Pak}},\ and\ \bibinfo {author} {\bibfnamefont {S.}~\bibnamefont
				{Hammes-Schiffer}},\ }\bibfield  {title} {\enquote {\bibinfo {title}
				{{Development of Electron-Proton Density Functionals for Multicomponent
						Density Functional Theory}},}\ }\href
		{https://doi.org/10.1103/PhysRevLett.101.153001} {\bibfield  {journal}
			{\bibinfo  {journal} {Phys. Rev. Lett.}\ }\textbf {\bibinfo {volume} {101}},\
			\bibinfo {pages} {153001} (\bibinfo {year} {2008})}\BibitemShut {NoStop}%
		\bibitem [{\citenamefont {Goings}, \citenamefont {Lestrange},\ and\
			\citenamefont {Li}(2018)}]{Goings2018}%
		\BibitemOpen
		\bibfield  {author} {\bibinfo {author} {\bibfnamefont {J.~J.}\ \bibnamefont
				{Goings}}, \bibinfo {author} {\bibfnamefont {P.~J.}\ \bibnamefont
				{Lestrange}},\ and\ \bibinfo {author} {\bibfnamefont {X.}~\bibnamefont
				{Li}},\ }\bibfield  {title} {\enquote {\bibinfo {title} {{Real-time
						time-dependent electronic structure theory}},}\ }\href
		{https://doi.org/10.1002/WCMS.1341} {\bibfield  {journal} {\bibinfo
				{journal} {Wiley Interdiscip. Rev.: Comput. Mol. Sci.}\ }\textbf {\bibinfo
				{volume} {8}},\ \bibinfo {pages} {e1341} (\bibinfo {year}
			{2018})}\BibitemShut {NoStop}%
		\bibitem [{\citenamefont {Isborn}\ and\ \citenamefont {Li}(2008)}]{Isborn2008}%
		\BibitemOpen
		\bibfield  {author} {\bibinfo {author} {\bibfnamefont {C.~M.}\ \bibnamefont
				{Isborn}}\ and\ \bibinfo {author} {\bibfnamefont {X.}~\bibnamefont {Li}},\
		}\bibfield  {title} {\enquote {\bibinfo {title} {{Modeling the doubly excited
						state with time-dependent Hartree–Fock and density functional theories}},}\
		}\href {https://doi.org/10.1063/1.3020336} {\bibfield  {journal} {\bibinfo
				{journal} {J. Chem. Phys.}\ }\textbf {\bibinfo {volume} {129}},\ \bibinfo
			{pages} {204107} (\bibinfo {year} {2008})}\BibitemShut {NoStop}%
		\bibitem [{\citenamefont {Galego}\ \emph {et~al.}(2019)\citenamefont {Galego},
			\citenamefont {Climent}, \citenamefont {Garcia-Vidal},\ and\ \citenamefont
			{Feist}}]{Galego2019}%
		\BibitemOpen
		\bibfield  {author} {\bibinfo {author} {\bibfnamefont {J.}~\bibnamefont
				{Galego}}, \bibinfo {author} {\bibfnamefont {C.}~\bibnamefont {Climent}},
			\bibinfo {author} {\bibfnamefont {F.~J.}\ \bibnamefont {Garcia-Vidal}},\ and\
			\bibinfo {author} {\bibfnamefont {J.}~\bibnamefont {Feist}},\ }\bibfield
		{title} {\enquote {\bibinfo {title} {{Cavity Casimir-Polder Forces and Their
						Effects in Ground-State Chemical Reactivity}},}\ }\href
		{https://doi.org/10.1103/PhysRevX.9.021057} {\bibfield  {journal} {\bibinfo
				{journal} {Phys. Rev. X}\ }\textbf {\bibinfo {volume} {9}},\ \bibinfo {pages}
			{021057} (\bibinfo {year} {2019})}\BibitemShut {NoStop}%
		\bibitem [{\citenamefont {Campos-Gonzalez-Angulo}, \citenamefont {Ribeiro},\
			and\ \citenamefont {Yuen-Zhou}(2019)}]{Campos-Gonzalez-Angulo2019}%
		\BibitemOpen
		\bibfield  {author} {\bibinfo {author} {\bibfnamefont {J.~A.}\ \bibnamefont
				{Campos-Gonzalez-Angulo}}, \bibinfo {author} {\bibfnamefont {R.~F.}\
				\bibnamefont {Ribeiro}},\ and\ \bibinfo {author} {\bibfnamefont
				{J.}~\bibnamefont {Yuen-Zhou}},\ }\bibfield  {title} {\enquote {\bibinfo
				{title} {{Resonant catalysis of thermally activated chemical reactions with
						vibrational polaritons}},}\ }\href
		{https://doi.org/10.1038/s41467-019-12636-1} {\bibfield  {journal} {\bibinfo
				{journal} {Nat. Commun.}\ }\textbf {\bibinfo {volume} {10}},\ \bibinfo
			{pages} {4685} (\bibinfo {year} {2019})}\BibitemShut {NoStop}%
		\bibitem [{\citenamefont {Li}, \citenamefont {Mandal},\ and\ \citenamefont
			{Huo}(2021)}]{LiHuo2021}%
		\BibitemOpen
		\bibfield  {author} {\bibinfo {author} {\bibfnamefont {X.}~\bibnamefont
				{Li}}, \bibinfo {author} {\bibfnamefont {A.}~\bibnamefont {Mandal}},\ and\
			\bibinfo {author} {\bibfnamefont {P.}~\bibnamefont {Huo}},\ }\bibfield
		{title} {\enquote {\bibinfo {title} {{Cavity Frequency-Dependent Theory for
						Vibrational Polariton Chemistry}},}\ }\href
		{https://doi.org/10.1038/s41467-021-21610-9} {\bibfield  {journal} {\bibinfo
				{journal} {Nat. Commun.}\ }\textbf {\bibinfo {volume} {12}},\ \bibinfo
			{pages} {1315} (\bibinfo {year} {2021})}\BibitemShut {NoStop}%
		\bibitem [{\citenamefont {Sch{\"{a}}fer}\ \emph {et~al.}(2022)\citenamefont
			{Sch{\"{a}}fer}, \citenamefont {Flick}, \citenamefont {Ronca}, \citenamefont
			{Narang},\ and\ \citenamefont {Rubio}}]{Schafer2021}%
		\BibitemOpen
		\bibfield  {author} {\bibinfo {author} {\bibfnamefont {C.}~\bibnamefont
				{Sch{\"{a}}fer}}, \bibinfo {author} {\bibfnamefont {J.}~\bibnamefont
				{Flick}}, \bibinfo {author} {\bibfnamefont {E.}~\bibnamefont {Ronca}},
			\bibinfo {author} {\bibfnamefont {P.}~\bibnamefont {Narang}},\ and\ \bibinfo
			{author} {\bibfnamefont {A.}~\bibnamefont {Rubio}},\ }\bibfield  {title}
		{\enquote {\bibinfo {title} {{Shining light on the microscopic resonant
						mechanism responsible for cavity-mediated chemical reactivity}},}\ }\href
		{https://doi.org/10.1038/s41467-022-35363-6} {\bibfield  {journal} {\bibinfo
				{journal} {Nat. Commun.}\ }\textbf {\bibinfo {volume} {13}},\ \bibinfo
			{pages} {7817} (\bibinfo {year} {2022})}\BibitemShut {NoStop}%
		\bibitem [{\citenamefont {Flick}\ \emph {et~al.}(2017)\citenamefont {Flick},
			\citenamefont {Ruggenthaler}, \citenamefont {Appel},\ and\ \citenamefont
			{Rubio}}]{Flick2017}%
		\BibitemOpen
		\bibfield  {author} {\bibinfo {author} {\bibfnamefont {J.}~\bibnamefont
				{Flick}}, \bibinfo {author} {\bibfnamefont {M.}~\bibnamefont {Ruggenthaler}},
			\bibinfo {author} {\bibfnamefont {H.}~\bibnamefont {Appel}},\ and\ \bibinfo
			{author} {\bibfnamefont {A.}~\bibnamefont {Rubio}},\ }\bibfield  {title}
		{\enquote {\bibinfo {title} {{Atoms and Molecules in Cavities, from Weak to
						Strong Coupling in Quantum-Electrodynamics (QED) Chemistry}},}\ }\href
		{https://doi.org/10.1073/pnas.1615509114} {\bibfield  {journal} {\bibinfo
				{journal} {Proc. Natl. Acad. Sci.}\ }\textbf {\bibinfo {volume} {114}},\
			\bibinfo {pages} {3026--3034} (\bibinfo {year} {2017})}\BibitemShut {NoStop}%
		\bibitem [{\citenamefont {Rosenzweig}\ \emph {et~al.}(2022)\citenamefont
			{Rosenzweig}, \citenamefont {Hoffmann}, \citenamefont {Lacombe},\ and\
			\citenamefont {Maitra}}]{Rosenzweig2022}%
		\BibitemOpen
		\bibfield  {author} {\bibinfo {author} {\bibfnamefont {B.}~\bibnamefont
				{Rosenzweig}}, \bibinfo {author} {\bibfnamefont {N.~M.}\ \bibnamefont
				{Hoffmann}}, \bibinfo {author} {\bibfnamefont {L.}~\bibnamefont {Lacombe}},\
			and\ \bibinfo {author} {\bibfnamefont {N.~T.}\ \bibnamefont {Maitra}},\
		}\bibfield  {title} {\enquote {\bibinfo {title} {{Analysis of the Classical
						Trajectory Treatment of Photon Dynamics for Polaritonic Phenomena}},}\ }\href
		{https://doi.org/10.1063/5.0079379/5.0079379.MM.ORIGINAL.V2.MP4} {\bibfield
			{journal} {\bibinfo  {journal} {J. Chem. Phys.}\ }\textbf {\bibinfo {volume}
				{156}},\ \bibinfo {pages} {054101} (\bibinfo {year} {2022})}\BibitemShut
		{NoStop}%
		\bibitem [{\citenamefont {Riso}\ \emph {et~al.}(2022)\citenamefont {Riso},
			\citenamefont {Haugland}, \citenamefont {Ronca},\ and\ \citenamefont
			{Koch}}]{Riso2022}%
		\BibitemOpen
		\bibfield  {author} {\bibinfo {author} {\bibfnamefont {R.~R.}\ \bibnamefont
				{Riso}}, \bibinfo {author} {\bibfnamefont {T.~S.}\ \bibnamefont {Haugland}},
			\bibinfo {author} {\bibfnamefont {E.}~\bibnamefont {Ronca}},\ and\ \bibinfo
			{author} {\bibfnamefont {H.}~\bibnamefont {Koch}},\ }\bibfield  {title}
		{\enquote {\bibinfo {title} {{Molecular Orbital Theory in Cavity QED
						Environments}},}\ }\href {https://doi.org/10.1038/s41467-022-29003-2}
		{\bibfield  {journal} {\bibinfo  {journal} {Nat. Commun.}\ }\textbf {\bibinfo
				{volume} {13}},\ \bibinfo {pages} {1368} (\bibinfo {year}
			{2022})}\BibitemShut {NoStop}%
		\bibitem [{\citenamefont {Epifanovsky}\ \emph {et~al.}(2021)\citenamefont
			{Epifanovsky}, \citenamefont {Gilbert}, \citenamefont {Feng}, \citenamefont
			{Lee}, \citenamefont {Mao}, \citenamefont {Mardirossian}, \citenamefont
			{Pokhilko}, \citenamefont {White}, \citenamefont {Coons}, \citenamefont
			{Dempwolff}, \citenamefont {Gan}, \citenamefont {Hait}, \citenamefont {Horn},
			\citenamefont {Jacobson}, \citenamefont {Kaliman}, \citenamefont {Kussmann},
			\citenamefont {Lange}, \citenamefont {Lao}, \citenamefont {Levine},
			\citenamefont {Liu}, \citenamefont {McKenzie}, \citenamefont {Morrison},
			\citenamefont {Nanda}, \citenamefont {Plasser}, \citenamefont {Rehn},
			\citenamefont {Vidal}, \citenamefont {You}, \citenamefont {Zhu},
			\citenamefont {Alam}, \citenamefont {Albrecht}, \citenamefont {Aldossary},
			\citenamefont {Alguire}, \citenamefont {Andersen}, \citenamefont {Athavale},
			\citenamefont {Barton}, \citenamefont {Begam}, \citenamefont {Behn},
			\citenamefont {Bellonzi}, \citenamefont {Bernard}, \citenamefont {Berquist},
			\citenamefont {Burton}, \citenamefont {Carreras}, \citenamefont
			{Carter-Fenk}, \citenamefont {Chakraborty}, \citenamefont {Chien},
			\citenamefont {Closser}, \citenamefont {Cofer-Shabica}, \citenamefont
			{Dasgupta}, \citenamefont {{De Wergifosse}}, \citenamefont {Deng},
			\citenamefont {Diedenhofen}, \citenamefont {Do}, \citenamefont {Ehlert},
			\citenamefont {Fang}, \citenamefont {Fatehi}, \citenamefont {Feng},
			\citenamefont {Friedhoff}, \citenamefont {Gayvert}, \citenamefont {Ge},
			\citenamefont {Gidofalvi}, \citenamefont {Goldey}, \citenamefont {Gomes},
			\citenamefont {Gonz{\'{a}}lez-Espinoza}, \citenamefont {Gulania},
			\citenamefont {Gunina}, \citenamefont {Hanson-Heine}, \citenamefont
			{Harbach}, \citenamefont {Hauser}, \citenamefont {Herbst}, \citenamefont
			{{Hern{\'{a}}ndez Vera}}, \citenamefont {Hodecker}, \citenamefont {Holden},
			\citenamefont {Houck}, \citenamefont {Huang}, \citenamefont {Hui},
			\citenamefont {Huynh}, \citenamefont {Ivanov}, \citenamefont {J{\'{a}}sz},
			\citenamefont {Ji}, \citenamefont {Jiang}, \citenamefont {Kaduk},
			\citenamefont {K{\"{a}}hler}, \citenamefont {Khistyaev}, \citenamefont {Kim},
			\citenamefont {Kis}, \citenamefont {Klunzinger}, \citenamefont
			{Koczor-Benda}, \citenamefont {Koh}, \citenamefont {Kosenkov}, \citenamefont
			{Koulias}, \citenamefont {Kowalczyk}, \citenamefont {Krauter}, \citenamefont
			{Kue}, \citenamefont {Kunitsa}, \citenamefont {Kus}, \citenamefont
			{Ladj{\'{a}}nszki}, \citenamefont {Landau}, \citenamefont {Lawler},
			\citenamefont {Lefrancois}, \citenamefont {Lehtola}, \citenamefont {Li},
			\citenamefont {Li}, \citenamefont {Liang}, \citenamefont {Liebenthal},
			\citenamefont {Lin}, \citenamefont {Lin}, \citenamefont {Liu}, \citenamefont
			{Liu}, \citenamefont {Loipersberger}, \citenamefont {Luenser}, \citenamefont
			{Manjanath}, \citenamefont {Manohar}, \citenamefont {Mansoor}, \citenamefont
			{Manzer}, \citenamefont {Mao}, \citenamefont {Marenich}, \citenamefont
			{Markovich}, \citenamefont {Mason}, \citenamefont {Maurer}, \citenamefont
			{McLaughlin}, \citenamefont {Menger}, \citenamefont {Mewes}, \citenamefont
			{Mewes}, \citenamefont {Morgante}, \citenamefont {Mullinax}, \citenamefont
			{Oosterbaan}, \citenamefont {Paran}, \citenamefont {Paul}, \citenamefont
			{Paul}, \citenamefont {Pavo{\v{s}}evi{\'{c}}}, \citenamefont {Pei},
			\citenamefont {Prager}, \citenamefont {Proynov}, \citenamefont {R{\'{a}}k},
			\citenamefont {Ramos-Cordoba}, \citenamefont {Rana}, \citenamefont {Rask},
			\citenamefont {Rettig}, \citenamefont {Richard}, \citenamefont {Rob},
			\citenamefont {Rossomme}, \citenamefont {Scheele}, \citenamefont {Scheurer},
			\citenamefont {Schneider}, \citenamefont {Sergueev}, \citenamefont {Sharada},
			\citenamefont {Skomorowski}, \citenamefont {Small}, \citenamefont {Stein},
			\citenamefont {Su}, \citenamefont {Sundstrom}, \citenamefont {Tao},
			\citenamefont {Thirman}, \citenamefont {Tornai}, \citenamefont {Tsuchimochi},
			\citenamefont {Tubman}, \citenamefont {Veccham}, \citenamefont {Vydrov},
			\citenamefont {Wenzel}, \citenamefont {Witte}, \citenamefont {Yamada},
			\citenamefont {Yao}, \citenamefont {Yeganeh}, \citenamefont {Yost},
			\citenamefont {Zech}, \citenamefont {Zhang}, \citenamefont {Zhang},
			\citenamefont {Zhang}, \citenamefont {Zuev}, \citenamefont {Aspuru-Guzik},
			\citenamefont {Bell}, \citenamefont {Besley}, \citenamefont {Bravaya},
			\citenamefont {Brooks}, \citenamefont {Casanova}, \citenamefont {Chai},
			\citenamefont {Coriani}, \citenamefont {Cramer}, \citenamefont {Cserey},
			\citenamefont {Deprince}, \citenamefont {Distasio}, \citenamefont {Dreuw},
			\citenamefont {Dunietz}, \citenamefont {Furlani}, \citenamefont {Goddard},
			\citenamefont {Hammes-Schiffer}, \citenamefont {Head-Gordon}, \citenamefont
			{Hehre}, \citenamefont {Hsu}, \citenamefont {Jagau}, \citenamefont {Jung},
			\citenamefont {Klamt}, \citenamefont {Kong}, \citenamefont {Lambrecht},
			\citenamefont {Liang}, \citenamefont {Mayhall}, \citenamefont {McCurdy},
			\citenamefont {Neaton}, \citenamefont {Ochsenfeld}, \citenamefont {Parkhill},
			\citenamefont {Peverati}, \citenamefont {Rassolov}, \citenamefont {Shao},
			\citenamefont {Slipchenko}, \citenamefont {Stauch}, \citenamefont {Steele},
			\citenamefont {Subotnik}, \citenamefont {Thom}, \citenamefont {Tkatchenko},
			\citenamefont {Truhlar}, \citenamefont {{Van Voorhis}}, \citenamefont
			{Wesolowski}, \citenamefont {Whaley}, \citenamefont {Woodcock}, \citenamefont
			{Zimmerman}, \citenamefont {Faraji}, \citenamefont {Gill}, \citenamefont
			{Head-Gordon}, \citenamefont {Herbert},\ and\ \citenamefont
			{Krylov}}]{Epifanovsky2021}%
		\BibitemOpen
		\bibfield  {author} {\bibinfo {author} {\bibfnamefont {E.}~\bibnamefont
				{Epifanovsky}}, \bibinfo {author} {\bibfnamefont {A.~T.}\ \bibnamefont
				{Gilbert}}, \bibinfo {author} {\bibfnamefont {X.}~\bibnamefont {Feng}},
			\bibinfo {author} {\bibfnamefont {J.}~\bibnamefont {Lee}}, \bibinfo {author}
			{\bibfnamefont {Y.}~\bibnamefont {Mao}}, \bibinfo {author} {\bibfnamefont
				{N.}~\bibnamefont {Mardirossian}}, \bibinfo {author} {\bibfnamefont
				{P.}~\bibnamefont {Pokhilko}}, \bibinfo {author} {\bibfnamefont {A.~F.}\
				\bibnamefont {White}}, \bibinfo {author} {\bibfnamefont {M.~P.}\ \bibnamefont
				{Coons}}, \bibinfo {author} {\bibfnamefont {A.~L.}\ \bibnamefont
				{Dempwolff}}, \bibinfo {author} {\bibfnamefont {Z.}~\bibnamefont {Gan}},
			\bibinfo {author} {\bibfnamefont {D.}~\bibnamefont {Hait}}, \bibinfo {author}
			{\bibfnamefont {P.~R.}\ \bibnamefont {Horn}}, \bibinfo {author}
			{\bibfnamefont {L.~D.}\ \bibnamefont {Jacobson}}, \bibinfo {author}
			{\bibfnamefont {I.}~\bibnamefont {Kaliman}}, \bibinfo {author} {\bibfnamefont
				{J.}~\bibnamefont {Kussmann}}, \bibinfo {author} {\bibfnamefont {A.~W.}\
				\bibnamefont {Lange}}, \bibinfo {author} {\bibfnamefont {K.~U.}\ \bibnamefont
				{Lao}}, \bibinfo {author} {\bibfnamefont {D.~S.}\ \bibnamefont {Levine}},
			\bibinfo {author} {\bibfnamefont {J.}~\bibnamefont {Liu}}, \bibinfo {author}
			{\bibfnamefont {S.~C.}\ \bibnamefont {McKenzie}}, \bibinfo {author}
			{\bibfnamefont {A.~F.}\ \bibnamefont {Morrison}}, \bibinfo {author}
			{\bibfnamefont {K.~D.}\ \bibnamefont {Nanda}}, \bibinfo {author}
			{\bibfnamefont {F.}~\bibnamefont {Plasser}}, \bibinfo {author} {\bibfnamefont
				{D.~R.}\ \bibnamefont {Rehn}}, \bibinfo {author} {\bibfnamefont {M.~L.}\
				\bibnamefont {Vidal}}, \bibinfo {author} {\bibfnamefont {Z.~Q.}\ \bibnamefont
				{You}}, \bibinfo {author} {\bibfnamefont {Y.}~\bibnamefont {Zhu}}, \bibinfo
			{author} {\bibfnamefont {B.}~\bibnamefont {Alam}}, \bibinfo {author}
			{\bibfnamefont {B.~J.}\ \bibnamefont {Albrecht}}, \bibinfo {author}
			{\bibfnamefont {A.}~\bibnamefont {Aldossary}}, \bibinfo {author}
			{\bibfnamefont {E.}~\bibnamefont {Alguire}}, \bibinfo {author} {\bibfnamefont
				{J.~H.}\ \bibnamefont {Andersen}}, \bibinfo {author} {\bibfnamefont
				{V.}~\bibnamefont {Athavale}}, \bibinfo {author} {\bibfnamefont
				{D.}~\bibnamefont {Barton}}, \bibinfo {author} {\bibfnamefont
				{K.}~\bibnamefont {Begam}}, \bibinfo {author} {\bibfnamefont
				{A.}~\bibnamefont {Behn}}, \bibinfo {author} {\bibfnamefont {N.}~\bibnamefont
				{Bellonzi}}, \bibinfo {author} {\bibfnamefont {Y.~A.}\ \bibnamefont
				{Bernard}}, \bibinfo {author} {\bibfnamefont {E.~J.}\ \bibnamefont
				{Berquist}}, \bibinfo {author} {\bibfnamefont {H.~G.}\ \bibnamefont
				{Burton}}, \bibinfo {author} {\bibfnamefont {A.}~\bibnamefont {Carreras}},
			\bibinfo {author} {\bibfnamefont {K.}~\bibnamefont {Carter-Fenk}}, \bibinfo
			{author} {\bibfnamefont {R.}~\bibnamefont {Chakraborty}}, \bibinfo {author}
			{\bibfnamefont {A.~D.}\ \bibnamefont {Chien}}, \bibinfo {author}
			{\bibfnamefont {K.~D.}\ \bibnamefont {Closser}}, \bibinfo {author}
			{\bibfnamefont {V.}~\bibnamefont {Cofer-Shabica}}, \bibinfo {author}
			{\bibfnamefont {S.}~\bibnamefont {Dasgupta}}, \bibinfo {author}
			{\bibfnamefont {M.}~\bibnamefont {{De Wergifosse}}}, \bibinfo {author}
			{\bibfnamefont {J.}~\bibnamefont {Deng}}, \bibinfo {author} {\bibfnamefont
				{M.}~\bibnamefont {Diedenhofen}}, \bibinfo {author} {\bibfnamefont
				{H.}~\bibnamefont {Do}}, \bibinfo {author} {\bibfnamefont {S.}~\bibnamefont
				{Ehlert}}, \bibinfo {author} {\bibfnamefont {P.~T.}\ \bibnamefont {Fang}},
			\bibinfo {author} {\bibfnamefont {S.}~\bibnamefont {Fatehi}}, \bibinfo
			{author} {\bibfnamefont {Q.}~\bibnamefont {Feng}}, \bibinfo {author}
			{\bibfnamefont {T.}~\bibnamefont {Friedhoff}}, \bibinfo {author}
			{\bibfnamefont {J.}~\bibnamefont {Gayvert}}, \bibinfo {author} {\bibfnamefont
				{Q.}~\bibnamefont {Ge}}, \bibinfo {author} {\bibfnamefont {G.}~\bibnamefont
				{Gidofalvi}}, \bibinfo {author} {\bibfnamefont {M.}~\bibnamefont {Goldey}},
			\bibinfo {author} {\bibfnamefont {J.}~\bibnamefont {Gomes}}, \bibinfo
			{author} {\bibfnamefont {C.~E.}\ \bibnamefont {Gonz{\'{a}}lez-Espinoza}},
			\bibinfo {author} {\bibfnamefont {S.}~\bibnamefont {Gulania}}, \bibinfo
			{author} {\bibfnamefont {A.~O.}\ \bibnamefont {Gunina}}, \bibinfo {author}
			{\bibfnamefont {M.~W.}\ \bibnamefont {Hanson-Heine}}, \bibinfo {author}
			{\bibfnamefont {P.~H.}\ \bibnamefont {Harbach}}, \bibinfo {author}
			{\bibfnamefont {A.}~\bibnamefont {Hauser}}, \bibinfo {author} {\bibfnamefont
				{M.~F.}\ \bibnamefont {Herbst}}, \bibinfo {author} {\bibfnamefont
				{M.}~\bibnamefont {{Hern{\'{a}}ndez Vera}}}, \bibinfo {author} {\bibfnamefont
				{M.}~\bibnamefont {Hodecker}}, \bibinfo {author} {\bibfnamefont {Z.~C.}\
				\bibnamefont {Holden}}, \bibinfo {author} {\bibfnamefont {S.}~\bibnamefont
				{Houck}}, \bibinfo {author} {\bibfnamefont {X.}~\bibnamefont {Huang}},
			\bibinfo {author} {\bibfnamefont {K.}~\bibnamefont {Hui}}, \bibinfo {author}
			{\bibfnamefont {B.~C.}\ \bibnamefont {Huynh}}, \bibinfo {author}
			{\bibfnamefont {M.}~\bibnamefont {Ivanov}}, \bibinfo {author} {\bibfnamefont
				{{\'{A}}.}~\bibnamefont {J{\'{a}}sz}}, \bibinfo {author} {\bibfnamefont
				{H.}~\bibnamefont {Ji}}, \bibinfo {author} {\bibfnamefont {H.}~\bibnamefont
				{Jiang}}, \bibinfo {author} {\bibfnamefont {B.}~\bibnamefont {Kaduk}},
			\bibinfo {author} {\bibfnamefont {S.}~\bibnamefont {K{\"{a}}hler}}, \bibinfo
			{author} {\bibfnamefont {K.}~\bibnamefont {Khistyaev}}, \bibinfo {author}
			{\bibfnamefont {J.}~\bibnamefont {Kim}}, \bibinfo {author} {\bibfnamefont
				{G.}~\bibnamefont {Kis}}, \bibinfo {author} {\bibfnamefont {P.}~\bibnamefont
				{Klunzinger}}, \bibinfo {author} {\bibfnamefont {Z.}~\bibnamefont
				{Koczor-Benda}}, \bibinfo {author} {\bibfnamefont {J.~H.}\ \bibnamefont
				{Koh}}, \bibinfo {author} {\bibfnamefont {D.}~\bibnamefont {Kosenkov}},
			\bibinfo {author} {\bibfnamefont {L.}~\bibnamefont {Koulias}}, \bibinfo
			{author} {\bibfnamefont {T.}~\bibnamefont {Kowalczyk}}, \bibinfo {author}
			{\bibfnamefont {C.~M.}\ \bibnamefont {Krauter}}, \bibinfo {author}
			{\bibfnamefont {K.}~\bibnamefont {Kue}}, \bibinfo {author} {\bibfnamefont
				{A.}~\bibnamefont {Kunitsa}}, \bibinfo {author} {\bibfnamefont
				{T.}~\bibnamefont {Kus}}, \bibinfo {author} {\bibfnamefont {I.}~\bibnamefont
				{Ladj{\'{a}}nszki}}, \bibinfo {author} {\bibfnamefont {A.}~\bibnamefont
				{Landau}}, \bibinfo {author} {\bibfnamefont {K.~V.}\ \bibnamefont {Lawler}},
			\bibinfo {author} {\bibfnamefont {D.}~\bibnamefont {Lefrancois}}, \bibinfo
			{author} {\bibfnamefont {S.}~\bibnamefont {Lehtola}}, \bibinfo {author}
			{\bibfnamefont {R.~R.}\ \bibnamefont {Li}}, \bibinfo {author} {\bibfnamefont
				{Y.~P.}\ \bibnamefont {Li}}, \bibinfo {author} {\bibfnamefont
				{J.}~\bibnamefont {Liang}}, \bibinfo {author} {\bibfnamefont
				{M.}~\bibnamefont {Liebenthal}}, \bibinfo {author} {\bibfnamefont {H.~H.}\
				\bibnamefont {Lin}}, \bibinfo {author} {\bibfnamefont {Y.~S.}\ \bibnamefont
				{Lin}}, \bibinfo {author} {\bibfnamefont {F.}~\bibnamefont {Liu}}, \bibinfo
			{author} {\bibfnamefont {K.~Y.}\ \bibnamefont {Liu}}, \bibinfo {author}
			{\bibfnamefont {M.}~\bibnamefont {Loipersberger}}, \bibinfo {author}
			{\bibfnamefont {A.}~\bibnamefont {Luenser}}, \bibinfo {author} {\bibfnamefont
				{A.}~\bibnamefont {Manjanath}}, \bibinfo {author} {\bibfnamefont
				{P.}~\bibnamefont {Manohar}}, \bibinfo {author} {\bibfnamefont
				{E.}~\bibnamefont {Mansoor}}, \bibinfo {author} {\bibfnamefont {S.~F.}\
				\bibnamefont {Manzer}}, \bibinfo {author} {\bibfnamefont {S.~P.}\
				\bibnamefont {Mao}}, \bibinfo {author} {\bibfnamefont {A.~V.}\ \bibnamefont
				{Marenich}}, \bibinfo {author} {\bibfnamefont {T.}~\bibnamefont {Markovich}},
			\bibinfo {author} {\bibfnamefont {S.}~\bibnamefont {Mason}}, \bibinfo
			{author} {\bibfnamefont {S.~A.}\ \bibnamefont {Maurer}}, \bibinfo {author}
			{\bibfnamefont {P.~F.}\ \bibnamefont {McLaughlin}}, \bibinfo {author}
			{\bibfnamefont {M.~F.}\ \bibnamefont {Menger}}, \bibinfo {author}
			{\bibfnamefont {J.~M.}\ \bibnamefont {Mewes}}, \bibinfo {author}
			{\bibfnamefont {S.~A.}\ \bibnamefont {Mewes}}, \bibinfo {author}
			{\bibfnamefont {P.}~\bibnamefont {Morgante}}, \bibinfo {author}
			{\bibfnamefont {J.~W.}\ \bibnamefont {Mullinax}}, \bibinfo {author}
			{\bibfnamefont {K.~J.}\ \bibnamefont {Oosterbaan}}, \bibinfo {author}
			{\bibfnamefont {G.}~\bibnamefont {Paran}}, \bibinfo {author} {\bibfnamefont
				{A.~C.}\ \bibnamefont {Paul}}, \bibinfo {author} {\bibfnamefont {S.~K.}\
				\bibnamefont {Paul}}, \bibinfo {author} {\bibfnamefont {F.}~\bibnamefont
				{Pavo{\v{s}}evi{\'{c}}}}, \bibinfo {author} {\bibfnamefont {Z.}~\bibnamefont
				{Pei}}, \bibinfo {author} {\bibfnamefont {S.}~\bibnamefont {Prager}},
			\bibinfo {author} {\bibfnamefont {E.~I.}\ \bibnamefont {Proynov}}, \bibinfo
			{author} {\bibfnamefont {{\'{A}}.}~\bibnamefont {R{\'{a}}k}}, \bibinfo
			{author} {\bibfnamefont {E.}~\bibnamefont {Ramos-Cordoba}}, \bibinfo {author}
			{\bibfnamefont {B.}~\bibnamefont {Rana}}, \bibinfo {author} {\bibfnamefont
				{A.~E.}\ \bibnamefont {Rask}}, \bibinfo {author} {\bibfnamefont
				{A.}~\bibnamefont {Rettig}}, \bibinfo {author} {\bibfnamefont {R.~M.}\
				\bibnamefont {Richard}}, \bibinfo {author} {\bibfnamefont {F.}~\bibnamefont
				{Rob}}, \bibinfo {author} {\bibfnamefont {E.}~\bibnamefont {Rossomme}},
			\bibinfo {author} {\bibfnamefont {T.}~\bibnamefont {Scheele}}, \bibinfo
			{author} {\bibfnamefont {M.}~\bibnamefont {Scheurer}}, \bibinfo {author}
			{\bibfnamefont {M.}~\bibnamefont {Schneider}}, \bibinfo {author}
			{\bibfnamefont {N.}~\bibnamefont {Sergueev}}, \bibinfo {author}
			{\bibfnamefont {S.~M.}\ \bibnamefont {Sharada}}, \bibinfo {author}
			{\bibfnamefont {W.}~\bibnamefont {Skomorowski}}, \bibinfo {author}
			{\bibfnamefont {D.~W.}\ \bibnamefont {Small}}, \bibinfo {author}
			{\bibfnamefont {C.~J.}\ \bibnamefont {Stein}}, \bibinfo {author}
			{\bibfnamefont {Y.~C.}\ \bibnamefont {Su}}, \bibinfo {author} {\bibfnamefont
				{E.~J.}\ \bibnamefont {Sundstrom}}, \bibinfo {author} {\bibfnamefont
				{Z.}~\bibnamefont {Tao}}, \bibinfo {author} {\bibfnamefont {J.}~\bibnamefont
				{Thirman}}, \bibinfo {author} {\bibfnamefont {G.~J.}\ \bibnamefont {Tornai}},
			\bibinfo {author} {\bibfnamefont {T.}~\bibnamefont {Tsuchimochi}}, \bibinfo
			{author} {\bibfnamefont {N.~M.}\ \bibnamefont {Tubman}}, \bibinfo {author}
			{\bibfnamefont {S.~P.}\ \bibnamefont {Veccham}}, \bibinfo {author}
			{\bibfnamefont {O.}~\bibnamefont {Vydrov}}, \bibinfo {author} {\bibfnamefont
				{J.}~\bibnamefont {Wenzel}}, \bibinfo {author} {\bibfnamefont
				{J.}~\bibnamefont {Witte}}, \bibinfo {author} {\bibfnamefont
				{A.}~\bibnamefont {Yamada}}, \bibinfo {author} {\bibfnamefont
				{K.}~\bibnamefont {Yao}}, \bibinfo {author} {\bibfnamefont {S.}~\bibnamefont
				{Yeganeh}}, \bibinfo {author} {\bibfnamefont {S.~R.}\ \bibnamefont {Yost}},
			\bibinfo {author} {\bibfnamefont {A.}~\bibnamefont {Zech}}, \bibinfo {author}
			{\bibfnamefont {I.~Y.}\ \bibnamefont {Zhang}}, \bibinfo {author}
			{\bibfnamefont {X.}~\bibnamefont {Zhang}}, \bibinfo {author} {\bibfnamefont
				{Y.}~\bibnamefont {Zhang}}, \bibinfo {author} {\bibfnamefont
				{D.}~\bibnamefont {Zuev}}, \bibinfo {author} {\bibfnamefont {A.}~\bibnamefont
				{Aspuru-Guzik}}, \bibinfo {author} {\bibfnamefont {A.~T.}\ \bibnamefont
				{Bell}}, \bibinfo {author} {\bibfnamefont {N.~A.}\ \bibnamefont {Besley}},
			\bibinfo {author} {\bibfnamefont {K.~B.}\ \bibnamefont {Bravaya}}, \bibinfo
			{author} {\bibfnamefont {B.~R.}\ \bibnamefont {Brooks}}, \bibinfo {author}
			{\bibfnamefont {D.}~\bibnamefont {Casanova}}, \bibinfo {author}
			{\bibfnamefont {J.~D.}\ \bibnamefont {Chai}}, \bibinfo {author}
			{\bibfnamefont {S.}~\bibnamefont {Coriani}}, \bibinfo {author} {\bibfnamefont
				{C.~J.}\ \bibnamefont {Cramer}}, \bibinfo {author} {\bibfnamefont
				{G.}~\bibnamefont {Cserey}}, \bibinfo {author} {\bibfnamefont {A.~E.}\
				\bibnamefont {Deprince}}, \bibinfo {author} {\bibfnamefont {R.~A.}\
				\bibnamefont {Distasio}}, \bibinfo {author} {\bibfnamefont {A.}~\bibnamefont
				{Dreuw}}, \bibinfo {author} {\bibfnamefont {B.~D.}\ \bibnamefont {Dunietz}},
			\bibinfo {author} {\bibfnamefont {T.~R.}\ \bibnamefont {Furlani}}, \bibinfo
			{author} {\bibfnamefont {W.~A.}\ \bibnamefont {Goddard}}, \bibinfo {author}
			{\bibfnamefont {S.}~\bibnamefont {Hammes-Schiffer}}, \bibinfo {author}
			{\bibfnamefont {T.}~\bibnamefont {Head-Gordon}}, \bibinfo {author}
			{\bibfnamefont {W.~J.}\ \bibnamefont {Hehre}}, \bibinfo {author}
			{\bibfnamefont {C.~P.}\ \bibnamefont {Hsu}}, \bibinfo {author} {\bibfnamefont
				{T.~C.}\ \bibnamefont {Jagau}}, \bibinfo {author} {\bibfnamefont
				{Y.}~\bibnamefont {Jung}}, \bibinfo {author} {\bibfnamefont {A.}~\bibnamefont
				{Klamt}}, \bibinfo {author} {\bibfnamefont {J.}~\bibnamefont {Kong}},
			\bibinfo {author} {\bibfnamefont {D.~S.}\ \bibnamefont {Lambrecht}}, \bibinfo
			{author} {\bibfnamefont {W.}~\bibnamefont {Liang}}, \bibinfo {author}
			{\bibfnamefont {N.~J.}\ \bibnamefont {Mayhall}}, \bibinfo {author}
			{\bibfnamefont {C.~W.}\ \bibnamefont {McCurdy}}, \bibinfo {author}
			{\bibfnamefont {J.~B.}\ \bibnamefont {Neaton}}, \bibinfo {author}
			{\bibfnamefont {C.}~\bibnamefont {Ochsenfeld}}, \bibinfo {author}
			{\bibfnamefont {J.~A.}\ \bibnamefont {Parkhill}}, \bibinfo {author}
			{\bibfnamefont {R.}~\bibnamefont {Peverati}}, \bibinfo {author}
			{\bibfnamefont {V.~A.}\ \bibnamefont {Rassolov}}, \bibinfo {author}
			{\bibfnamefont {Y.}~\bibnamefont {Shao}}, \bibinfo {author} {\bibfnamefont
				{L.~V.}\ \bibnamefont {Slipchenko}}, \bibinfo {author} {\bibfnamefont
				{T.}~\bibnamefont {Stauch}}, \bibinfo {author} {\bibfnamefont {R.~P.}\
				\bibnamefont {Steele}}, \bibinfo {author} {\bibfnamefont {J.~E.}\
				\bibnamefont {Subotnik}}, \bibinfo {author} {\bibfnamefont {A.~J.}\
				\bibnamefont {Thom}}, \bibinfo {author} {\bibfnamefont {A.}~\bibnamefont
				{Tkatchenko}}, \bibinfo {author} {\bibfnamefont {D.~G.}\ \bibnamefont
				{Truhlar}}, \bibinfo {author} {\bibfnamefont {T.}~\bibnamefont {{Van
						Voorhis}}}, \bibinfo {author} {\bibfnamefont {T.~A.}\ \bibnamefont
				{Wesolowski}}, \bibinfo {author} {\bibfnamefont {K.~B.}\ \bibnamefont
				{Whaley}}, \bibinfo {author} {\bibfnamefont {H.~L.}\ \bibnamefont
				{Woodcock}}, \bibinfo {author} {\bibfnamefont {P.~M.}\ \bibnamefont
				{Zimmerman}}, \bibinfo {author} {\bibfnamefont {S.}~\bibnamefont {Faraji}},
			\bibinfo {author} {\bibfnamefont {P.~M.}\ \bibnamefont {Gill}}, \bibinfo
			{author} {\bibfnamefont {M.}~\bibnamefont {Head-Gordon}}, \bibinfo {author}
			{\bibfnamefont {J.~M.}\ \bibnamefont {Herbert}},\ and\ \bibinfo {author}
			{\bibfnamefont {A.~I.}\ \bibnamefont {Krylov}},\ }\bibfield  {title}
		{\enquote {\bibinfo {title} {{Software for the Frontiers of Quantum
						Chemistry: An Overview of Developments in the Q-Chem 5 Package}},}\ }\href
		{https://doi.org/10.1063/5.0055522} {\bibfield  {journal} {\bibinfo
				{journal} {J. Chem. Phys.}\ }\textbf {\bibinfo {volume} {155}},\ \bibinfo
			{pages} {084801} (\bibinfo {year} {2021})}\BibitemShut {NoStop}%
		\bibitem [{\citenamefont {Li}\ \emph {et~al.}(2005{\natexlab{b}})\citenamefont
			{Li}, \citenamefont {Smith}, \citenamefont {Markevitch}, \citenamefont
			{Romanov}, \citenamefont {Levis},\ and\ \citenamefont {Schlegel}}]{Li2005}%
		\BibitemOpen
		\bibfield  {author} {\bibinfo {author} {\bibfnamefont {X.}~\bibnamefont
				{Li}}, \bibinfo {author} {\bibfnamefont {S.~M.}\ \bibnamefont {Smith}},
			\bibinfo {author} {\bibfnamefont {A.~N.}\ \bibnamefont {Markevitch}},
			\bibinfo {author} {\bibfnamefont {D.~A.}\ \bibnamefont {Romanov}}, \bibinfo
			{author} {\bibfnamefont {R.~J.}\ \bibnamefont {Levis}},\ and\ \bibinfo
			{author} {\bibfnamefont {H.~B.}\ \bibnamefont {Schlegel}},\ }\bibfield
		{title} {\enquote {\bibinfo {title} {{A Time-Dependent Hartree--Fock Approach
						for Studying the Electronic Optical Response of Molecules in Intense
						Fields}},}\ }\href {https://doi.org/10.1039/B415849K} {\bibfield  {journal}
			{\bibinfo  {journal} {Phys. Chem. Chem. Phys.}\ }\textbf {\bibinfo {volume}
				{7}},\ \bibinfo {pages} {233--239} (\bibinfo {year}
			{2005}{\natexlab{b}})}\BibitemShut {NoStop}%
		\bibitem [{\citenamefont {{De Santis}}\ \emph {et~al.}(2020)\citenamefont {{De
					Santis}}, \citenamefont {Storchi}, \citenamefont {Belpassi}, \citenamefont
			{Quiney},\ and\ \citenamefont {Tarantelli}}]{DeSantis2020}%
		\BibitemOpen
		\bibfield  {author} {\bibinfo {author} {\bibfnamefont {M.}~\bibnamefont {{De
						Santis}}}, \bibinfo {author} {\bibfnamefont {L.}~\bibnamefont {Storchi}},
			\bibinfo {author} {\bibfnamefont {L.}~\bibnamefont {Belpassi}}, \bibinfo
			{author} {\bibfnamefont {H.~M.}\ \bibnamefont {Quiney}},\ and\ \bibinfo
			{author} {\bibfnamefont {F.}~\bibnamefont {Tarantelli}},\ }\bibfield  {title}
		{\enquote {\bibinfo {title} {{PyBERTHART: A Relativistic Real-Time
						Four-Component TDDFT Implementation Using Prototyping Techniques Based on
						Python.}}}\ }\href {https://doi.org/10.1021/acs.jctc.0c00053} {\bibfield
			{journal} {\bibinfo  {journal} {J. Chem. Theory Comput.}\ }\textbf {\bibinfo
				{volume} {16}},\ \bibinfo {pages} {2410--2429} (\bibinfo {year}
			{2020})}\BibitemShut {NoStop}%
		\bibitem [{\citenamefont {Lee}, \citenamefont {Yang},\ and\ \citenamefont
			{Parr}(1988)}]{Lee1988}%
		\BibitemOpen
		\bibfield  {author} {\bibinfo {author} {\bibfnamefont {C.}~\bibnamefont
				{Lee}}, \bibinfo {author} {\bibfnamefont {W.}~\bibnamefont {Yang}},\ and\
			\bibinfo {author} {\bibfnamefont {R.~G.}\ \bibnamefont {Parr}},\ }\bibfield
		{title} {\enquote {\bibinfo {title} {{Development of the Colle--Salvetti
						Correlation-Energy Formula into a Functional of the Electron Density}},}\
		}\href {https://doi.org/10.1103/PhysRevB.37.785} {\bibfield  {journal}
			{\bibinfo  {journal} {Phys. Rev. B}\ }\textbf {\bibinfo {volume} {37}},\
			\bibinfo {pages} {785} (\bibinfo {year} {1988})}\BibitemShut {NoStop}%
		\bibitem [{\citenamefont {Becke}(1988)}]{Becke1988}%
		\BibitemOpen
		\bibfield  {author} {\bibinfo {author} {\bibfnamefont {A.~D.}\ \bibnamefont
				{Becke}},\ }\bibfield  {title} {\enquote {\bibinfo {title}
				{{Density-Functional Exchange-Energy Approximation with Correct Asymptotic
						Behavior}},}\ }\href {https://doi.org/10.1103/PhysRevA.38.3098} {\bibfield
			{journal} {\bibinfo  {journal} {Phys. Rev. A}\ }\textbf {\bibinfo {volume}
				{38}},\ \bibinfo {pages} {3098} (\bibinfo {year} {1988})}\BibitemShut
		{NoStop}%
		\bibitem [{\citenamefont {Becke}(1998)}]{Becke1998}%
		\BibitemOpen
		\bibfield  {author} {\bibinfo {author} {\bibfnamefont {A.~D.}\ \bibnamefont
				{Becke}},\ }\bibfield  {title} {\enquote {\bibinfo {title} {{A New
						Inhomogeneity Parameter in Density-Functional Theory}},}\ }\href
		{https://doi.org/10.1063/1.476722} {\bibfield  {journal} {\bibinfo  {journal}
				{J. Chem. Phys.}\ }\textbf {\bibinfo {volume} {109}},\ \bibinfo {pages}
			{2092} (\bibinfo {year} {1998})}\BibitemShut {NoStop}%
		\bibitem [{\citenamefont {Brorsen}, \citenamefont {Yang},\ and\ \citenamefont
			{Hammes-Schiffer}(2017)}]{Brorsen2017}%
		\BibitemOpen
		\bibfield  {author} {\bibinfo {author} {\bibfnamefont {K.~R.}\ \bibnamefont
				{Brorsen}}, \bibinfo {author} {\bibfnamefont {Y.}~\bibnamefont {Yang}},\ and\
			\bibinfo {author} {\bibfnamefont {S.}~\bibnamefont {Hammes-Schiffer}},\
		}\bibfield  {title} {\enquote {\bibinfo {title} {{Multicomponent Density
						Functional Theory: Impact of Nuclear Quantum Effects on Proton Affinities and
						Geometries}},}\ }\href {https://doi.org/10.1021/acs.jpclett.7b01442}
		{\bibfield  {journal} {\bibinfo  {journal} {J.Phys. Chem. Lett.}\ }\textbf
			{\bibinfo {volume} {8}},\ \bibinfo {pages} {3488--3493} (\bibinfo {year}
			{2017})}\BibitemShut {NoStop}%
		\bibitem [{\citenamefont {Yang}\ \emph {et~al.}(2017)\citenamefont {Yang},
			\citenamefont {Brorsen}, \citenamefont {Culpitt}, \citenamefont {Pak},\ and\
			\citenamefont {Hammes-Schiffer}}]{Yang2017}%
		\BibitemOpen
		\bibfield  {author} {\bibinfo {author} {\bibfnamefont {Y.}~\bibnamefont
				{Yang}}, \bibinfo {author} {\bibfnamefont {K.~R.}\ \bibnamefont {Brorsen}},
			\bibinfo {author} {\bibfnamefont {T.}~\bibnamefont {Culpitt}}, \bibinfo
			{author} {\bibfnamefont {M.~V.}\ \bibnamefont {Pak}},\ and\ \bibinfo {author}
			{\bibfnamefont {S.}~\bibnamefont {Hammes-Schiffer}},\ }\bibfield  {title}
		{\enquote {\bibinfo {title} {{Development of A Practical Multicomponent
						Density Functional for Electron-Proton Correlation to Produce Accurate Proton
						Densities}},}\ }\href {https://doi.org/10.1063/1.4996038} {\bibfield
			{journal} {\bibinfo  {journal} {J. Chem. Phys.}\ }\textbf {\bibinfo {volume}
				{147}},\ \bibinfo {pages} {114113} (\bibinfo {year} {2017})}\BibitemShut
		{NoStop}%
		\bibitem [{\citenamefont {Dunning}(1989)}]{Dunning1989}%
		\BibitemOpen
		\bibfield  {author} {\bibinfo {author} {\bibfnamefont {T.~H.}\ \bibnamefont
				{Dunning}},\ }\bibfield  {title} {\enquote {\bibinfo {title} {{Gaussian Basis
						Sets for Use in Correlated Molecular Calculations. I. The Atoms Boron Through
						Neon and Hydrogen}},}\ }\href {https://doi.org/10.1063/1.456153} {\bibfield
			{journal} {\bibinfo  {journal} {J. Chem. Phys.}\ }\textbf {\bibinfo {volume}
				{90}},\ \bibinfo {pages} {1007--1023} (\bibinfo {year} {1989})}\BibitemShut
		{NoStop}%
		\bibitem [{\citenamefont {Yu}, \citenamefont {Pavo{\v{s}}evi{\'{c}}},\ and\
			\citenamefont {Hammes-Schiffer}(2020)}]{Yu2020NEOBasis}%
		\BibitemOpen
		\bibfield  {author} {\bibinfo {author} {\bibfnamefont {Q.}~\bibnamefont
				{Yu}}, \bibinfo {author} {\bibfnamefont {F.}~\bibnamefont
				{Pavo{\v{s}}evi{\'{c}}}},\ and\ \bibinfo {author} {\bibfnamefont
				{S.}~\bibnamefont {Hammes-Schiffer}},\ }\bibfield  {title} {\enquote
			{\bibinfo {title} {{Development of nuclear basis sets for multicomponent
						quantum chemistry methods}},}\ }\href {https://doi.org/10.1063/5.0009233}
		{\bibfield  {journal} {\bibinfo  {journal} {J. Chem. Phys.}\ }\textbf
			{\bibinfo {volume} {152}},\ \bibinfo {pages} {244123} (\bibinfo {year}
			{2020})}\BibitemShut {NoStop}%
		\bibitem [{\citenamefont {Bruner}, \citenamefont {Lamaster},\ and\
			\citenamefont {Lopata}(2016)}]{Bruner2016}%
		\BibitemOpen
		\bibfield  {author} {\bibinfo {author} {\bibfnamefont {A.}~\bibnamefont
				{Bruner}}, \bibinfo {author} {\bibfnamefont {D.}~\bibnamefont {Lamaster}},\
			and\ \bibinfo {author} {\bibfnamefont {K.}~\bibnamefont {Lopata}},\
		}\bibfield  {title} {\enquote {\bibinfo {title} {{Accelerated Broadband
						Spectra Using Transition Dipole Decomposition and Pad{\'{e}}
						Approximants}},}\ }\href
		{https://doi.org/10.1021/ACS.JCTC.6B00511/SUPPL_FILE/CT6B00511_SI_001.PDF}
		{\bibfield  {journal} {\bibinfo  {journal} {J. Chem. Theory Comput.}\
			}\textbf {\bibinfo {volume} {12}},\ \bibinfo {pages} {3741--3750} (\bibinfo
			{year} {2016})}\BibitemShut {NoStop}%
		\bibitem [{\citenamefont {Culpitt}\ \emph {et~al.}(2019)\citenamefont
			{Culpitt}, \citenamefont {Yang}, \citenamefont {Pavo{\v{s}}evi{\'{c}}},
			\citenamefont {Tao},\ and\ \citenamefont {Hammes-Schiffer}}]{Culpitt2019JCP}%
		\BibitemOpen
		\bibfield  {author} {\bibinfo {author} {\bibfnamefont {T.}~\bibnamefont
				{Culpitt}}, \bibinfo {author} {\bibfnamefont {Y.}~\bibnamefont {Yang}},
			\bibinfo {author} {\bibfnamefont {F.}~\bibnamefont {Pavo{\v{s}}evi{\'{c}}}},
			\bibinfo {author} {\bibfnamefont {Z.}~\bibnamefont {Tao}},\ and\ \bibinfo
			{author} {\bibfnamefont {S.}~\bibnamefont {Hammes-Schiffer}},\ }\bibfield
		{title} {\enquote {\bibinfo {title} {{Enhancing the Applicability of
						Multicomponent Time-Dependent Density Functional Theory}},}\ }\href
		{https://doi.org/10.1063/1.5099093} {\bibfield  {journal} {\bibinfo
				{journal} {J. Chem. Phys.}\ }\textbf {\bibinfo {volume} {150}},\ \bibinfo
			{pages} {201101} (\bibinfo {year} {2019})}\BibitemShut {NoStop}%
		\bibitem [{\citenamefont {Yang}\ \emph {et~al.}(2020)\citenamefont {Yang},
			\citenamefont {Pei}, \citenamefont {Deng}, \citenamefont {Mao}, \citenamefont
			{Wu}, \citenamefont {Yang}, \citenamefont {Wang}, \citenamefont {Aikens},
			\citenamefont {Liang},\ and\ \citenamefont {Shao}}]{Shao2020PCCP}%
		\BibitemOpen
		\bibfield  {author} {\bibinfo {author} {\bibfnamefont {J.}~\bibnamefont
				{Yang}}, \bibinfo {author} {\bibfnamefont {Z.}~\bibnamefont {Pei}}, \bibinfo
			{author} {\bibfnamefont {J.}~\bibnamefont {Deng}}, \bibinfo {author}
			{\bibfnamefont {Y.}~\bibnamefont {Mao}}, \bibinfo {author} {\bibfnamefont
				{Q.}~\bibnamefont {Wu}}, \bibinfo {author} {\bibfnamefont {Z.}~\bibnamefont
				{Yang}}, \bibinfo {author} {\bibfnamefont {B.}~\bibnamefont {Wang}}, \bibinfo
			{author} {\bibfnamefont {C.~M.}\ \bibnamefont {Aikens}}, \bibinfo {author}
			{\bibfnamefont {W.}~\bibnamefont {Liang}},\ and\ \bibinfo {author}
			{\bibfnamefont {Y.}~\bibnamefont {Shao}},\ }\bibfield  {title} {\enquote
			{\bibinfo {title} {Analysis and visualization of energy densities. i.
					insights from real-time time-dependent density functional theory
					simulations},}\ }\href {https://doi.org/10.1039/D0CP04206D} {\bibfield
			{journal} {\bibinfo  {journal} {Phys. Chem. Chem. Phys.}\ }\textbf {\bibinfo
				{volume} {22}},\ \bibinfo {pages} {26838--26851} (\bibinfo {year}
			{2020})}\BibitemShut {NoStop}%
		\bibitem [{\citenamefont {{Frisk Kockum}}\ \emph {et~al.}(2019)\citenamefont
			{{Frisk Kockum}}, \citenamefont {Miranowicz}, \citenamefont {{De Liberato}},
			\citenamefont {Savasta},\ and\ \citenamefont {Nori}}]{FriskKockum2019}%
		\BibitemOpen
		\bibfield  {author} {\bibinfo {author} {\bibfnamefont {A.}~\bibnamefont
				{{Frisk Kockum}}}, \bibinfo {author} {\bibfnamefont {A.}~\bibnamefont
				{Miranowicz}}, \bibinfo {author} {\bibfnamefont {S.}~\bibnamefont {{De
						Liberato}}}, \bibinfo {author} {\bibfnamefont {S.}~\bibnamefont {Savasta}},\
			and\ \bibinfo {author} {\bibfnamefont {F.}~\bibnamefont {Nori}},\ }\bibfield
		{title} {\enquote {\bibinfo {title} {{Ultrastrong coupling between light and
						matter}},}\ }\href {https://doi.org/10.1038/s42254-018-0006-2} {\bibfield
			{journal} {\bibinfo  {journal} {Nat. Rev. Phys.}\ }\textbf {\bibinfo {volume}
				{1}},\ \bibinfo {pages} {19--40} (\bibinfo {year} {2019})}\BibitemShut
		{NoStop}%
		\bibitem [{\citenamefont {Li}, \citenamefont {Subotnik},\ and\ \citenamefont
			{Nitzan}(2020)}]{Li2020Water}%
		\BibitemOpen
		\bibfield  {author} {\bibinfo {author} {\bibfnamefont {T.~E.}\ \bibnamefont
				{Li}}, \bibinfo {author} {\bibfnamefont {J.~E.}\ \bibnamefont {Subotnik}},\
			and\ \bibinfo {author} {\bibfnamefont {A.}~\bibnamefont {Nitzan}},\
		}\bibfield  {title} {\enquote {\bibinfo {title} {{Cavity Molecular Dynamics
						Simulations of Liquid Water under Vibrational Ultrastrong Coupling}},}\
		}\href {https://doi.org/10.1073/pnas.2009272117} {\bibfield  {journal}
			{\bibinfo  {journal} {Proc. Natl. Acad. Sci.}\ }\textbf {\bibinfo {volume}
				{117}},\ \bibinfo {pages} {18324--18331} (\bibinfo {year}
			{2020})}\BibitemShut {NoStop}%
		\bibitem [{\citenamefont {George}\ \emph {et~al.}(2015)\citenamefont {George},
			\citenamefont {Shalabney}, \citenamefont {Hutchison}, \citenamefont {Genet},\
			and\ \citenamefont {Ebbesen}}]{George2015}%
		\BibitemOpen
		\bibfield  {author} {\bibinfo {author} {\bibfnamefont {J.}~\bibnamefont
				{George}}, \bibinfo {author} {\bibfnamefont {A.}~\bibnamefont {Shalabney}},
			\bibinfo {author} {\bibfnamefont {J.~A.}\ \bibnamefont {Hutchison}}, \bibinfo
			{author} {\bibfnamefont {C.}~\bibnamefont {Genet}},\ and\ \bibinfo {author}
			{\bibfnamefont {T.~W.}\ \bibnamefont {Ebbesen}},\ }\bibfield  {title}
		{\enquote {\bibinfo {title} {{Liquid-Phase Vibrational Strong Coupling}},}\
		}\href {https://doi.org/10.1021/acs.jpclett.5b00204} {\bibfield  {journal}
			{\bibinfo  {journal} {J. Phys. Chem. Lett.}\ }\textbf {\bibinfo {volume}
				{6}},\ \bibinfo {pages} {1027--1031} (\bibinfo {year} {2015})}\BibitemShut
		{NoStop}%
		\bibitem [{\citenamefont {Yang}\ \emph {et~al.}(2021)\citenamefont {Yang},
			\citenamefont {Ou}, \citenamefont {Pei}, \citenamefont {Wang}, \citenamefont
			{Weng}, \citenamefont {Shuai}, \citenamefont {Mullen},\ and\ \citenamefont
			{Shao}}]{Yang2021QEDFT}%
		\BibitemOpen
		\bibfield  {author} {\bibinfo {author} {\bibfnamefont {J.}~\bibnamefont
				{Yang}}, \bibinfo {author} {\bibfnamefont {Q.}~\bibnamefont {Ou}}, \bibinfo
			{author} {\bibfnamefont {Z.}~\bibnamefont {Pei}}, \bibinfo {author}
			{\bibfnamefont {H.}~\bibnamefont {Wang}}, \bibinfo {author} {\bibfnamefont
				{B.}~\bibnamefont {Weng}}, \bibinfo {author} {\bibfnamefont {Z.}~\bibnamefont
				{Shuai}}, \bibinfo {author} {\bibfnamefont {K.}~\bibnamefont {Mullen}},\ and\
			\bibinfo {author} {\bibfnamefont {Y.}~\bibnamefont {Shao}},\ }\bibfield
		{title} {\enquote {\bibinfo {title} {{Quantum-Electrodynamical Time-Dependent
						Density Functional Theory within Gaussian Atomic Basis}},}\ }\href
		{https://doi.org/10.1063/5.0057542} {\bibfield  {journal} {\bibinfo
				{journal} {J. Chem. Phys.}\ }\textbf {\bibinfo {volume} {155}},\ \bibinfo
			{pages} {064107} (\bibinfo {year} {2021})}\BibitemShut {NoStop}%
		\bibitem [{\citenamefont {Luo}, \citenamefont {Fuks},\ and\ \citenamefont
			{Maitra}(2016)}]{Neepa2016JCP}%
		\BibitemOpen
		\bibfield  {author} {\bibinfo {author} {\bibfnamefont {K.}~\bibnamefont
				{Luo}}, \bibinfo {author} {\bibfnamefont {J.~I.}\ \bibnamefont {Fuks}},\ and\
			\bibinfo {author} {\bibfnamefont {N.~T.}\ \bibnamefont {Maitra}},\ }\bibfield
		{title} {\enquote {\bibinfo {title} {Studies of spuriously shifting
					resonances in time-dependent density functional theory},}\ }\href@noop {}
		{\bibfield  {journal} {\bibinfo  {journal} {J. Chem. Phys.}\ }\textbf
			{\bibinfo {volume} {145}},\ \bibinfo {pages} {044101} (\bibinfo {year}
			{2016})}\BibitemShut {NoStop}%
		\bibitem [{Note1()}]{Note1}%
		\BibitemOpen
		\bibinfo {note} {Note that a more general definition of proton transfer time
			would require the proton to form a stable bond with the acceptor
			oxygen.}\BibitemShut {Stop}%
		\bibitem [{\citenamefont {Hammes-Schiffer}(2015)}]{SHS2015PCET}%
		\BibitemOpen
		\bibfield  {author} {\bibinfo {author} {\bibfnamefont {S.}~\bibnamefont
				{Hammes-Schiffer}},\ }\bibfield  {title} {\enquote {\bibinfo {title}
				{{Proton-Coupled Electron Transfer: Moving Together and Charging Forward}},}\
		}\href {https://doi.org/10.1021/jacs.5b04087} {\bibfield  {journal} {\bibinfo
				{journal} {J. Am. Chem. Soc.}\ }\textbf {\bibinfo {volume} {137}},\ \bibinfo
			{pages} {8860--8871} (\bibinfo {year} {2015})}\BibitemShut {NoStop}%
	\end{thebibliography}
	
	%

	\clearpage
	\onecolumngrid
	

\section*{Supplementary material}

# Supplementary Material

# Electronic Born–Oppenheimer Approximation in Nuclear-Electronic Orbital Dynamics

Tao E. Li[a)] and Sharon Hammes-Schiffer[b)]

*Department of Chemistry, Yale University, New Haven, Connecticut 06520, USA*

---

[a)] Electronic mail: tao.li@yale.edu
[b)] Electronic mail: sharon.hammes-schiffer@yale.edu

# S1. INITIAL GEOMETRIES FOR HCN AND MALONALDEHYDE

HCN (Units: Angstrom)
N 0.0492158067 0.000 0.000
C 1.2046693425 0.000 0.000
H 2.1221148508 0.000 0.000

Malonaldehyde (Units: Angstrom; Gh: additional proton basis centers)
O 0.0000000000 -1.3008730953 2.0456624985
O 0.0000000000 1.2908329047 2.0456624985
C 0.0000000000 -1.2160310953 0.7592594985
C 0.0000000000 1.2059909047 0.7592594985
C 0.0000000000 -0.0050200953 0.0533444985
H 0.0000000000 -0.0050200953 -1.0248235015
H 0.0000000000 -2.1631460953 0.2189674985
H 0.0000000000 2.1531059047 0.2189674985
H 0.0000000000 -0.3010120550 2.3554201375
Gh 0.0000000000 0.0000000000 2.3554165853
Gh 0.0000000000 0.2553440000 2.3869010000

## S2. ADDITIONAL SIMULATION RESULTS

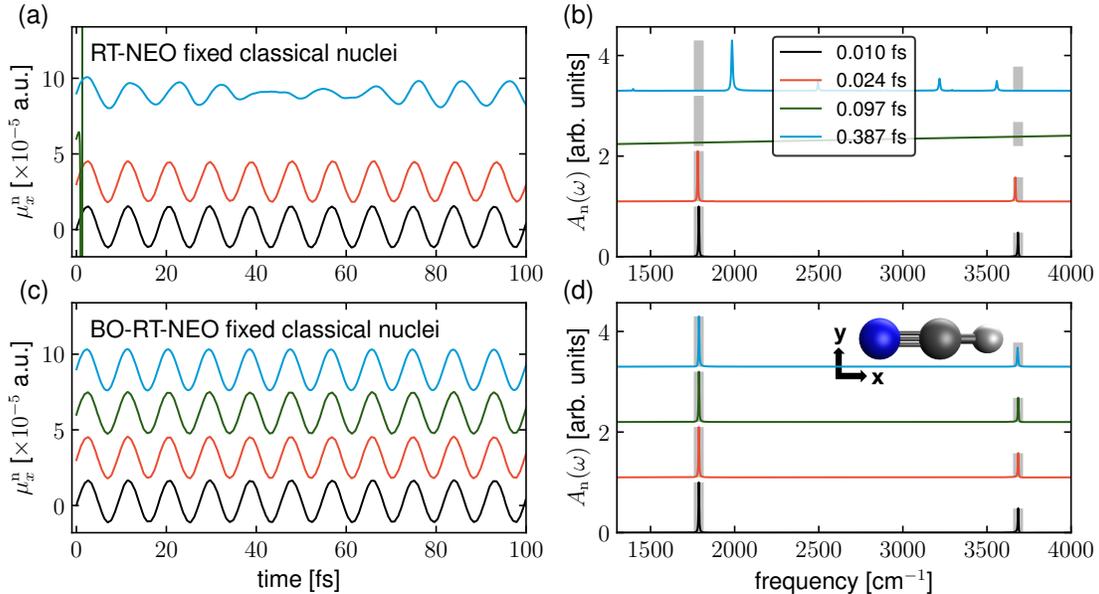

FIG. S1. The same plots as those shown in Fig. 1 except that here the absorption spectrum is plotted: $A_\mathrm{n} = \sum_{i=x,y,z} -\omega \mathrm{Im}\left[\mathcal{F}\left[\mu_i^\mathrm{n}(t)e^{-\gamma t}\right]\right]$.[S1] The linear-response peak heights (vertical gray solid lines) indicate the absorption intensities. Here, the peak ratios between the real-time signals and the linear-response signals (vertical gray lines) agree exactly.

TABLE S1. Linear-response NEO-TDDFT results for HCN.

| Excite state | Frequency | Electronic oscillator strength | Protonic oscillator strength |
|---|---|---|---|
| 1 | 1787 cm$^{-1}$ | $9.73 \times 10^{-5}$ a.u. | 0.3559 a.u. |
| 2 | 1787 cm$^{-1}$ | $9.73 \times 10^{-5}$ a.u. | 0.3559 a.u. |
| 3 | 3685 cm$^{-1}$ | $9.56 \times 10^{-5}$ a.u. | 0.3416 a.u. |

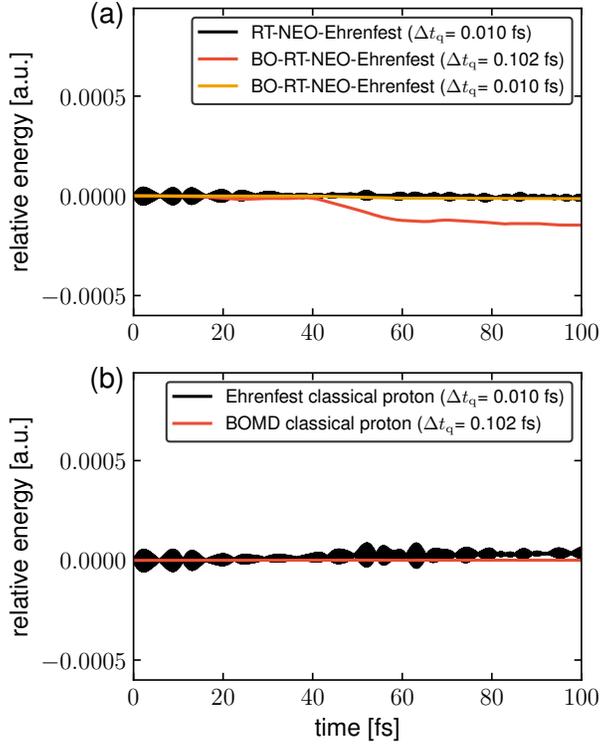

FIG. S2. Energy conservation along the trajectories shown in Fig. 3. For both the NEO and classical simulations, the black lines indicate the trajectories without the BO approximation (with a time step $\Delta t_\mathrm{q} = 0.010$ fs), and the red lines indicate the trajectories with the BO approximation (with a time step $\Delta t_\mathrm{q} = 0.102$ fs). The additional yellow line in part (a) indicates the BO-RT-NEO-Ehrenfest trajectory with the same time step as the RT-NEO-Ehrenfest dynamics (black line), showing that the energy conservation behavior of BO-RT-NEO-Ehrenfest dynamics can be improved by reducing the time step.

# S3. DETAILED ALGORITHMS FOR NEO DYNAMICS

**Algorithm S1** Semiclassical BO-RT-NEO dynamics for polaritons with fixed classical nuclei.

1: Calculate $\mu_\lambda(\tau = 0) = \text{Tr}\left[\mathbf{P}^{n'}(\tau = 0)\hat{\mu}_\lambda^{n'}\right] + 2\text{Tr}\left[\mathbf{P}^{e'}(\tau = 0)\hat{\mu}_\lambda^{e'}\right]$
2: **for** $\tau = \Delta t_q, 2\Delta t_q, \cdots$ **do**
3:     Calculate $\mu_\lambda(\tau) = \text{Tr}\left[\mathbf{P}^{n'}(\tau)\hat{\mu}_\lambda^{n'}\right] + 2\text{Tr}\left[\mathbf{P}^{e'}(\tau)\hat{\mu}_\lambda^{e'}\right] - \mu_\lambda(\tau = 0)$
4:     $p_{k,\lambda}(\tau + \frac{1}{2}\Delta t_q) = p_{k,\lambda}(\tau - \frac{1}{2}\Delta t_q) - [\omega_{k,\lambda}^2 q_{k,\lambda}(\tau) + \varepsilon_{k,\lambda}\mu_\lambda(\tau)]\Delta t_q$
5:     $q_{k,\lambda}(\tau + \Delta t_q) = q_{k,\lambda}(\tau) + p_{k,\lambda}(\tau + \frac{1}{2}\Delta t_q)\Delta t_q$
6:     $p_{k,\lambda}(\tau + \frac{1}{2}\Delta t_q) *= e^{-\gamma_c \Delta t_q}$   //cavity loss
7:     $\mathbf{P}^{e(n)'}(\tau) = [\mathbf{S}^{e(n)}]^{-1/2}\mathbf{P}^{e(n)}(\tau)[\mathbf{S}^{e(n)}]^{-1/2}$
8:     Build $\mathbf{F}_{\text{incav}}^{e(n)'}(\tau)$ using $\mathbf{P}^{e(n)'}(\tau)$ and $q_{k,\lambda}(\tau)$
9:     $\mathbf{F}_{\text{incav}}^{e(n)}(\tau + \frac{1}{2}\Delta t_q) = 2\mathbf{F}_{\text{incav}}^{e(n)}(\tau) - \mathbf{F}_{\text{incav}}^{e(n)}(\tau - \frac{1}{2}\Delta t_q)$
10:     counter = 1
11:     **while** True **do**
12:         $\mathbf{P}^n(\tau + \Delta t_q) = e^{-i\Delta t_q \mathbf{F}_{\text{incav}}^n(\tau + \frac{1}{2}\Delta t_q)}\mathbf{P}^n(\tau)e^{i\Delta t_q \mathbf{F}_{\text{incav}}^n(\tau + \frac{1}{2}\Delta t_q)}$
13:         **if** scf_e AND counter == 1 **then**
14:             Converge $\mathbf{P}^{e'}(\tau + \Delta t_q)$ to ground state given $\mathbf{P}^{n'}(\tau + \Delta t_q)$
15:         **else**
16:             $\mathbf{P}^e(\tau + \Delta t_q) = e^{-i\Delta t_q \mathbf{F}_{\text{incav}}^e(\tau + \frac{1}{2}\Delta t_q)}\mathbf{P}^e(\tau)e^{i\Delta t_q \mathbf{F}_{\text{incav}}^e(\tau + \frac{1}{2}\Delta t_q)}$
17:         **end if**
18:         Build $\mathbf{F}_{\text{incav}}^{e(n)'}(\tau + \Delta t_q)$ using $\mathbf{P}^{e(n)'}(\tau + \Delta t_q)$ and $q_{k,\lambda}(\tau + \Delta t_q)$
19:         $\mathbf{F}_{\text{incav}}^{e(n)}(\tau + \frac{1}{2}\Delta t_q) = \frac{1}{2}\mathbf{F}_{\text{incav}}^{e(n)}(\tau) + \frac{1}{2}\mathbf{F}_{\text{incav}}^{e(n)}(\tau + \Delta t_q)$
20:         **if** counter > 1 **then**
21:             **if** $|\mathbf{P}^{e(n)}(\tau + \Delta t_q) - \mathbf{P}_{\text{test}}^{e(n)}| <$ thres **then**
22:                 Exit the while loop
23:             **end if**
24:         **end if**
25:         $\mathbf{P}_{\text{test}}^{e(n)} = \mathbf{P}^{e(n)}(\tau + \Delta t_q)$
26:         counter += 1
27:     **end while**
28: **end for**

**Algorithm S2** BO-RT-NEO-Ehrenfest dynamics.
1: $\Delta t_{N_q} = \Delta t_N/n$, $\Delta t_q = \Delta t_{N_q}/m$
2: **for** $t = 0, \Delta t_N, 2\Delta t_N, \cdots$ **do**
3:     **if** `scf_e` **then**
4:         Converge $\mathbf{P}^{e'}(t)$ to ground state given $\mathbf{P}^{n'}(t)$
5:     **end if**
6:     Compute forces $\mathbf{F}(t)$ using $\mathbf{P}^{e'}(t)$ and $\mathbf{P}^{n'}(t)$
7:     $\mathbf{P}(t + \frac{1}{2}\Delta t_N) = \mathbf{P}(t - \frac{1}{2}\Delta t_N) + \mathbf{F}(t)\Delta t_N$
8:     $\mathbf{R}(t + \Delta t_N) = \mathbf{R}(t) + \mathbf{P}(t + \frac{1}{2}\Delta t_N)\Delta t_N/\mathbf{M}$
9:     **for** $j = 1, 2, \cdots, n$ **do**
10:         $t' = t + (j-1)\Delta t_{N_q}$
11:         $\mathbf{R}(t' + \frac{1}{2}\Delta t_{N_q}) = \mathbf{R}(t) + ((j-1)\Delta t_{N_q} + \frac{1}{2}\Delta t_{N_q})\mathbf{P}(t + \frac{1}{2}t_N)/\mathbf{M}$
12:         Update basis center to $\mathbf{R}(t' + \frac{1}{2}\Delta t_{N_q})$ and recompute $\mathbf{S}^{e(n)}$
13:         **for** $i = 1, 2, \cdots, m$ **do**
14:             $\tau = t + (i-1)\Delta t_q$
15:             $\mathbf{P}^{e(n)'}(\tau) = [\mathbf{S}^{e(n)}]^{-1/2}\mathbf{P}^{e(n)}(\tau)[\mathbf{S}^{e(n)}]^{-1/2}$
16:             Build $\mathbf{F}^{e(n)'}(\tau)$ using $\mathbf{P}^{e(n)'}(\tau)$ and $\mathbf{R}(t' + \frac{1}{2}\Delta t_{N_q})$
17:             $\mathbf{F}^{e(n)}(\tau + \frac{1}{2}\Delta t_q) = 2\mathbf{F}^{e(n)}(\tau) - \mathbf{F}^{e(n)}(\tau - \frac{1}{2}\Delta t_q)$
18:             `counter = 1`
19:             **while** True **do**
20:                 $\mathbf{P}^n(\tau + \Delta t_q) = e^{-i\Delta t_q[\mathbf{F}^n(\tau + \frac{1}{2}\Delta t_q)]}\mathbf{P}^n(\tau)e^{i\Delta t_q[\mathbf{F}^n(\tau + \frac{1}{2}\Delta t_q)]}$
21:                 **if** `scf_e` AND `counter == 1` **then**
22:                       Converge $\mathbf{P}^{e'}(\tau + \Delta t_q)$ to ground state given $\mathbf{P}^{n'}(\tau + \Delta t_q)$
23:                 **else**
24:                       $\mathbf{P}^e(\tau + \Delta t_q) = e^{-i\Delta t_q\mathbf{F}^e(\tau + \frac{1}{2}\Delta t_q)}\mathbf{P}^e(\tau)e^{i\Delta t_q\mathbf{F}^e(\tau + \frac{1}{2}\Delta t_q)}$
25:                 **end if**
26:                 Build $\mathbf{F}^{e(n)'}(\tau + \Delta t_q)$ using $\mathbf{P}^{e(n)'}(\tau + \Delta t_q)$ and $\mathbf{R}(t' + \frac{1}{2}\Delta t_{N_q})$
27:                 $\mathbf{F}^{e(n)}(\tau + \frac{1}{2}\Delta t_q) = \frac{1}{2}\mathbf{F}^{e(n)}(\tau) + \frac{1}{2}\mathbf{F}^{e(n)}(\tau + \Delta t_q)$
28:                 **if** `counter > 1` **then**
29:                       **if** $|\mathbf{P}^{e(n)}(\tau + \Delta t_q) - \mathbf{P}^{e(n)}_{\text{test}}| <$ `thres` **then**
30:                           Exit the while loop
31:                       **end if**
32:                 **end if**
33:                 $\mathbf{P}^{e(n)}_{\text{test}} = \mathbf{P}^{e(n)}(\tau + \Delta t_q)$
34:                 `counter += 1`
35:             **end while**
36:         **end for**
37:     **end for**
38: **end for**

In the above two algorithms, when the electronic BO approximation is invoked, the parameter `scf_e` is set to `true`, and otherwise this parameter is set to `false`.

The parameter `thres` in the above two algorithms controls the maximal error in the density matrices during each time step. In our simulation, we have set a very loose threshold

$10^{-4}$ a.u.

In RT-NEO-Ehrenfest dynamics, the parameters $m$ and $n$ are used to reduce the number of energy gradient evaluations,[S2] which is the most time-consuming step. For our calculations in the manuscript, we have set $m$ = 1 and $n$ = 10.

## A. Strategies for reducing the computational cost of BO-RT-NEO dynamics

For both algorithms above, one time consuming step is the building of the complex-valued electronic Fock matrix. Under the electronic Born–Oppenheimer (BO) approximation, building the complex-valued electronic Fock matrices in Lines 8 and 18 of Algorithm S1 and in Lines 16 and 26 of Algorithm S2 can be avoided. Instead, only the real-valued electronic Fock matrix needs to be built during the electronic SCF procedure. Although the electronic SCF procedure is time consuming, this strategy allows each BO-RT-NEO-TDDFT time step to be only twice the computational cost of a RT-NEO-TDDFT time step without the BO approximation.

For RT-NEO-Ehrenfest dynamics, the most time consuming step is the gradient evaluation. Under the electronic BO approximation, because the electronic density is always real-valued, it is unnecessary to calculate the gradient components involving the imaginary part of the electronic density. In contrast, for NEO-Ehrenfest dynamics without the electronic BO approximation, the imaginary part of the electronic density should be taken into account for the gradient evaluation. Hence, the computational cost of the gradient evaluation can also be reduced significantly under the electronic BO approximation.

## REFERENCES

[S1] L. Zhao, Z. Tao, F. Pavošević, A. Wildman, S. Hammes-Schiffer, and X. Li, "Real-Time Time-Dependent Nuclear-Electronic Orbital Approach: Dynamics beyond the Born-Oppenheimer Approximation," J. Phys. Chem. Lett. **11**, 4052–4058 (2020).

[S2] L. Zhao, A. Wildman, Z. Tao, P. Schneider, S. Hammes-Schiffer, and X. Li, "Nuclear–electronic orbital Ehrenfest dynamics," J. Chem. Phys. **153**, 224111 (2020).



\end{document}